\documentclass[12pt,onecolumn,article]{IEEEtran}
\usepackage{epsfig,url}
\usepackage{amsmath,graphicx,subcaption}
\usepackage{tikz,cite}
\usetikzlibrary{decorations.pathreplacing}
\usetikzlibrary{dsp,chains,arrows,shapes,spy,fit}
\usepackage{caption}
\usepackage{enumitem}
\tikzset{
    font={\fontsize{9}{11.0476pt}\selectfont}}
\usepackage{mathtools,amssymb,cuted,bm}
\usepackage{pgfplots}
\pgfplotsset{compat=newest}
\usepackage{romannum}
\usepackage{algpseudocode}
\usepackage{algorithm}
\begin{document}

\title{Joint Localization and Information Transfer for Reconfigurable Intelligent Surface Aided Full-Duplex Systems}

\author{Zhichao~Shao,~\IEEEmembership{Member,~IEEE},~Xiaojun~Yuan,~\IEEEmembership{Senior~Member,~IEEE},~Wei~Zhang,~\IEEEmembership{Fellow,~IEEE},~and~Marco~Di~Renzo,~\IEEEmembership{Fellow,~IEEE}
	\thanks{Z. Shao and X. Yuan are with the National Key Laboratory of Science and Technology on Communications, University of Electronic Science and Technology of China, Chengdu 611731, China (e-mail: {zhichao.shao; xjyuan}@uestc.edu.cn).}
	\thanks{W. Zhang is  with the School of Electrical Engineering and
		Telecommunications, University of New South Wales, Sydney, NSW 2052,
		Australia (e-mail: w.zhang@unsw.edu.au).}
	\thanks{M. Di Renzo is with Universit\'{e} Paris-Saclay, CNRS, CentraleSup\'{e}lec,
		Laboratoire des Signaux et Systèmes, 3 Rue Joliot-Curie, 91192 Gif-sur-Yvette, France. (marco.di-renzo@universite-paris-saclay.fr)}}

\maketitle

\begin{abstract}
In this work, we investigate a reconfigurable intelligent surface (RIS) aided integrated sensing and communication scenario, where a base station (BS) communicates with multiple devices in a full-duplex mode, and senses the positions of these devices simultaneously. An RIS is assumed to be mounted on each device to enhance the reflected echoes. Meanwhile, the information of each device is passively transferred to the BS via reflection modulation. We aim to tackle the problem of joint localization and information retrieval at the BS. A grid based parametric model is constructed and the joint estimation problem is formulated as a compressive sensing problem. We propose a novel message-passing algorithm to solve the considered problem, and a progressive approximation method to reduce the computational complexity involved in the message passing. Moreover, an expectation-maximization (EM) algorithm is applied for tuning the grid parameters to mitigate the model mismatch problem. Finally, we analyze the efficacy of the proposed algorithm through the Bayesian Cram\'er-Rao bound. Numerical results demonstrate the feasibility of the proposed scheme and the superior performance of the proposed EM-based message-passing algorithm.
\end{abstract}


\section{Introduction}
\IEEEPARstart{R}{econfigurable} intelligent surface (RIS), also known as intelligent reflecting surface, or large intelligent surface, has been recognized as a promising next-generation wireless communication technology \cite{8466374,9140329,9424177}. An RIS consists of a large number of software-controlled meta-atoms, each of which can independently impose the required phase shift on the incident electromagnetic waves. By carefully adjusting the phase shifts of all the meta-atoms, the radiated electromagnetic waves can be shaped to propagate towards desired directions, thereby significantly enhancing the communication quality. In wireless communications, RISs have many potential applications. For example, in RIS-enhanced cellular networks, RISs are deployed to establish favorable non-line-of-sight links between base stations (BSs) and users. Thus,
additional degrees of freedom are introduced by RISs for improving the system performance \cite{8741198}. Other applications include the beamforming design in simultaneous wireless information and power transfer networks \cite{8941080,9133435}, unmanned aerial vehicle aided wireless networks \cite{8959174}, and internet of things networks \cite{9148610}. 

Moreover, an RIS, as a large passive scattering array, generally has remarkable capabilities to sense the environments, which has motivated research activities towards the integration of sensing and communications (ISAC) in RIS aided systems. The main objective of an ISAC system is to enable coexistence between the communication and sensing functionalities by ensuring reliable communications with the users while using the same spectrum for sensing and localizing targets \cite{9737357,9376324,8999605}. The authors of \cite{9364358} employed one RIS for both sensing and communication where the direct path between the dual function radar and communication BS and the target exists. The goal was to maximize the radar signal-to-interference-plus-noise ratio (SINR) under the communication SINR constraint. The authors of \cite{sankar2021joint} proposed an RIS that is adaptively partitioned into two parts for communication and localization, respectively, when no direct path exists.

Motivated by the above discussions, in this work we investigate the RIS assisted communication system by designing a configuration with integrated sensing capability. The main advantage of the considered system is that with the aid of current communication network the sensing can be introduced by adding dedicated sensing processing from received communication signals, which averts the extra deployment of sensory networks. One typical example is the perceptive mobile network \cite{9296833,9349171}, which evolves from the current communication-only mobile network. It is expected to serve as a ubiquitous radar-sensing network, whilst providing uncompromising mobile communication services.
Specifically, we propose a new RIS-aided ISAC scenario, where a BS not only communicates with multiple devices (such as vehicles) in full-duplex, but also senses the positions of these devices simultaneously. We assume that an RIS is mounted on each device to enhance the echoes reflected by the device. Meanwhile, the information of each device is passively transferred to the BS via reflection modulation \cite{8941126,9117136}, in which the information is encoded into the phase adjustments of RIS meta-atoms. \emph{The proposed ISAC scenario is advantageous in two aspects. First, the communication system is full-duplex in frequency and time, where the information delivered by the BS can be received by the devices using conventional receivers, and, at the same time/frequency slot, the information of the devices is passively delivered to the BS via reflection modulation thanks to the RISs. Second, the devices are “green” since they do not emit any electromagnetic signals during the whole process.}

We consider a MIMO orthogonal frequency division multiplexing (OFDM) full-duplex ISAC system, and focus on the receiver design of the BS to jointly locating the positions of the devices and retrieving the information passively transferred by the RISs. OFDM technique, which transmits data symbols over orthogonal subcarriers, is popular for implementing ISAC systems. The authors of \cite{8169087} allocated different subcarriers to the communication and sensing functions, but this reduces the available bandwidth of each function. In \cite{7833233}, a high-resolution compressed sensing (CS) algorithm is proposed to solve the problem of joint delay-Doppler estimation of moving targets under the assumption that the signal is sparse in the Fourier-domain, but the computational cost is high while solving the problem of semidefinite program. The authors of \cite{7164629} derived a low-complexity subspace-based algorithm by applying a smoothing approach for joint estimation of range and Doppler shift in OFDM-based radar system, but the estimation accuracy for multiple targets is not high enough. 

The main contributions of this work are summarized as follows.
\begin{itemize}
	\item We establish a grid-based parametric system model, and formulate the joint estimation problem as a CS problem by exploiting the sparsity on the considered parameter grid.
	\item Based on a factor graph representation of the probability model, we develop a novel message-passing algorithm to efficiently solve the considered problem. During the message passing, a progressive approximation method is introduced to reduce the computational complexity.
	\item An expectation-maximization (EM) algorithm is applied for tuning the grid parameters and mitigating the model mismatch problem.
	\item The Bayesian Cram\'er-Rao bound (BCRB) is derived as a fundamental performance limit to evaluate the efficacy of the proposed algorithm.
	\item Numerical results demonstrate the feasibility of the proposed scenario, as well as the superior performance of the proposed EM-learning method.
\end{itemize}


The rest of this paper is organized as follows: in Section
\Romannum{2} the system model for the RIS-aided full-duplex MIMO-OFDM system is described, and the passive beamforming and information transfer of RIS is illustrated based on the generalized Snell's law and the reflection modulation, separately. In Section \Romannum{3} we propose the grid-based parametric model and solve the the localization and information recovery problem through our proposed message-passing algorithm. In Section \Romannum{4}, an EM-based parameters learning algorithm is presented for estimating prior parameters (i.e., noise variance and sparsity) for the proposed algorithm,  and the grid parameters for the possible model mismatch problem. In Section \Romannum{5}, we derive the analytical BCRB for the proposed system. Simulations are presented and discussed in Section \Romannum{6} and the paper is concluded in Section \Romannum{7}.

\emph{Notation}: Throughout the paper, bold letters indicate vectors and
matrices, non-bold letters express scalars if not particularly indicated. The operators $(\cdot)^T$, $(\cdot)^H$ and $(\cdot)^{-1}$ stand for the
transposition, Hermitian transposition and inverse,
respectively. $\mathbf{I}_n$ denotes a $n\times n$ identity matrix and
$\mathbf{0}_{m\times n}$ is an $m\times n$ all zeros matrix.
Additionally, $\text{diag}(\mathbf{a})$ is a diagonal matrix, where the diagonal elements are the vector of $\mathbf{a}$. $\otimes$, $Tr\{\cdot\}$ and $\text{vec}\{\cdot\}$ are denoted by the operation of Kronecker product, trace and vectorization, respectively. $Re\{\cdot\}$, $Im\{\cdot\}$ and $\angle$ extract the real parts, imaginary parts and phases from the complex values, respectively. Finally, $E\{\cdot\}$ and $Var\{\cdot\}$ calculate the expectation and the variance, respectively. $\mathcal{O}(\cdot)$ is the big O notation.

\section{System Description}\label{sec_2}
\subsection{RIS Aided Full-Duplex System}
\label{sec2_a}
We consider a MIMO-OFDM full-duplex system, where a BS with $N_\mathrm{t}$ transmit and $N_\mathrm{r}$ receive antennas communicates with multiple devices, each equipped with an RIS with $L$ reflecting elements and a conventional receiver. The antenna/element spacing is $\frac{\lambda}{2}$, where $\lambda$ is the wavelength.
\begin{figure}
	\centering
	\includegraphics[width=0.55\columnwidth]{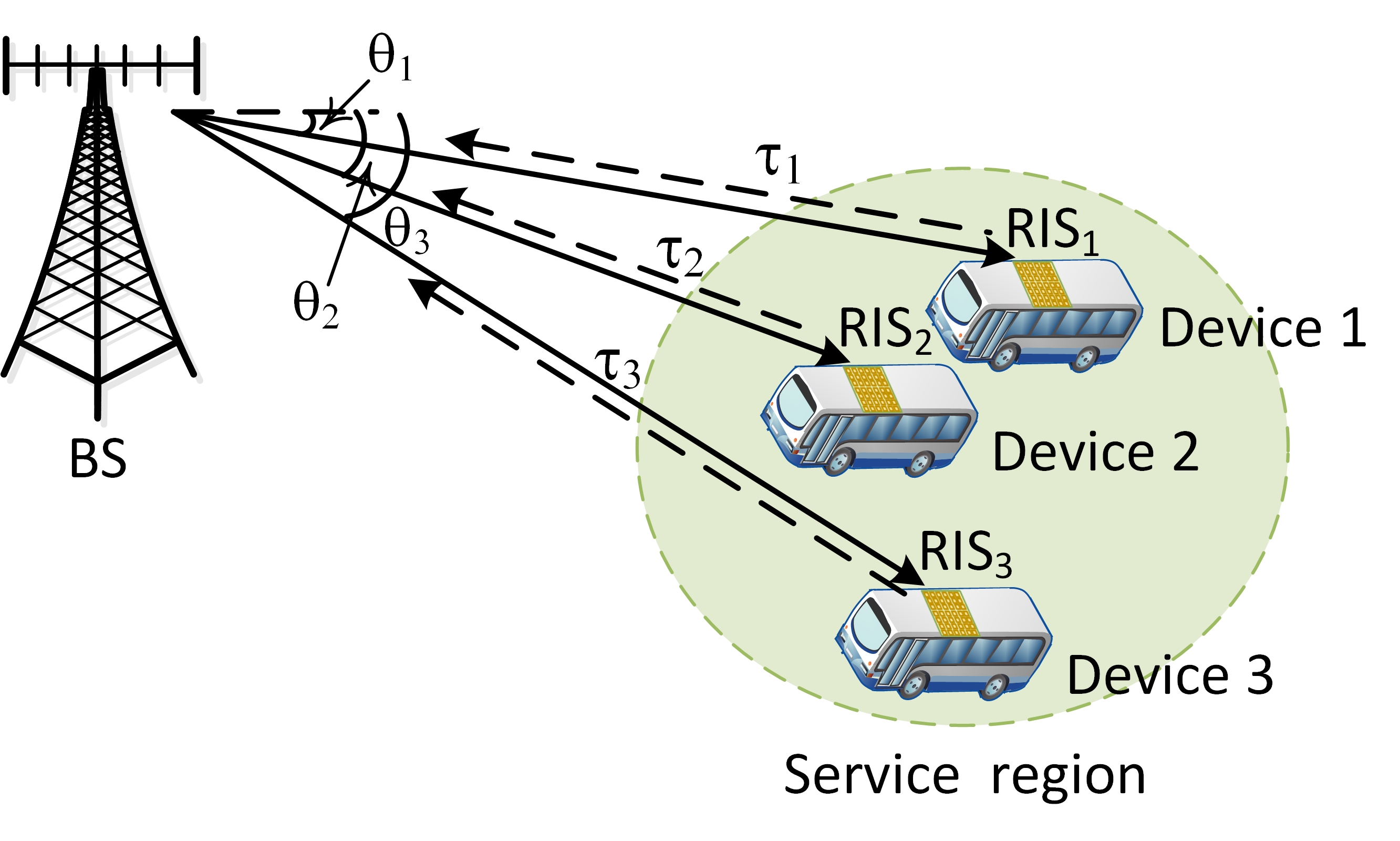}
	\caption{An RIS-aided full-duplex system.}
	\label{fig_model}
\end{figure}
The OFDM system has $N$ orthogonal subcarriers, and the frequency spacing of adjacent subcarriers is $\Delta f = 1/T$, where $T$ denotes the duration of one OFDM symbol. At the beginning of each OFDM symbol, a cyclic prefix (CP) of length $T_ \mathrm{cp}$ is inserted to avoid the intersymbol interference, and the duration of each OFDM block is $T_ \mathrm{b}=T+T_\mathrm{cp}$. 

The transmission protocol in Fig. \ref{fig_model} is described as follows. The BS first broadcasts the OFDM signal towards a given region via beamforming. The devices in the considered region receive the transmitted signal and recover the message from the BS by using traditional signal processing techniques\footnote{We omit the details here since the receiver design at the device is standard and is not the focus of this paper.}. Meanwhile, the RIS deployed on each device applies reflection modulation onto the incident electromagnetic wave, and reflects it back to the BS. Unlike the incident wave, the reflected wave contains new information that is sent from the RIS to the BS. Finally, the BS jointly estimates the positions and retrieves the information of the devices from the received echos.

The transmitted signal in the $m$-th OFDM block at the BS is given by 
\begin{equation}
		\mathbf{x}_{m}(t) = \sum_{n=0}^{N-1}\mathbf{x}_{m}[n]e^{j2\pi n\Delta ft}\Xi(t-mT_ \mathrm{b}),\quad mT_ \mathrm{b}\leq t\leq (m+1)T_ \mathrm{b},
	\label{equ_xn}
\end{equation}
where $\mathbf{x}_{m}[n]=[x_{m,1}[n],\cdots,x_{m,N_\mathrm{t}}[n]]^T\in\mathbb{C}^{N_\mathrm{t}\times1}$ contains the transmitted symbols at the $n$-th subcarrier in the $m$-th block; and
\begin{equation}
	\nonumber
	\Xi(t)=
	\begin{cases}
		1, & t\in[0,T_\mathrm{b}],\\
		0, & \text{otherwise}.
	\end{cases}
\end{equation}
We assume that the line of sight path exists and dominates the channel between the BS and each device. In fact, the considered application is different from conventional RIS-aided networks, where in the latter an RIS is utilized to create an additional reliable channel path. In this work, the RISs are utilized to generate backscattered signals for localization and data transmission in a passive manner. We model the signal reflected by each device as a multipath characterized by an angle of propagation and a delay. Considering that there are $K$ devices in the service region, the delay of the $k$-th path is denoted by $\tau_k$. We assume that the localization parameters remain constant in one transmission frame consisting of $M$ OFDM blocks, i.e. the angles and delays do not change in the period of one transmission frame. This is reasonable, since the duration of each transmission frame is typically in millisecond, and the distance traveled by the moving device can be ignored in such a short period of time. Furthermore, since the BS transmits and receives signals at the same time-frequency slot, the interference between the transmit and receive antennas needs to be considered while modeling the echos from the devices. In this work, we assume that the receive antennas are carefully isolated from the transmit antennas, so that the self-interference can be ignored \cite{6702851}. The echos received at the BS are contaminated by an additive noise $\mathbf{w}_m(t)\in\mathbb{C}^{N_\mathrm{r}\times 1}$ as
\begin{equation}\label{equ_ymt}
	\mathbf{y}_m(t)=\sum_{k=0}^{K-1}\beta_k\mathbf{H}_{\mathrm{UL},k}\mathbf{\Lambda}_{k,m}\mathbf{H}_{\mathrm{DL},k}\mathbf{x}_m(t-\tau_k)+\mathbf{w}_m(t),
\end{equation}   
where $\beta_k\in\mathbb{C}^{1\times 1}$ is the round-trip fading coefficient accounting for the path attenuation from the BS to the $k$-th device and from the $k$-th device back to the BS; $\mathbf{\Lambda}_{k,m}=\text{diag}\{[e^{j\theta_{k,m,1,1}},\cdots,e^{j\theta_{k,m,N_\mathrm{x},N_\mathrm{y}}}]\}\in\mathbb{C}^{L\times L}$ is a diagonal matrix whose diagonal elements are the phase shifts $\{\theta_{k,m,i,j}\}$ applied by the $k$-th RIS in the $m$-th OFDM block; $\mathbf{H}_{\mathrm{UL},k}\in\mathbb{C}^{N_\mathrm{r}\times L}$ and $\mathbf{H}_{\mathrm{DL},k}\in\mathbb{C}^{L\times N_\mathrm{t}}$ are the uplink and downlink channel matrices between the BS and the $k$-th device, respectively, given by
\begin{equation}
	\mathbf{H}_{\mathrm{DL},k}=\mathbf{a}_\mathrm{RIS}(\phi_{\mathrm{i},k},\gamma_{\mathrm{i},k})\mathbf{a}_\mathrm{BS}(\vartheta_k)^H
	\label{equ_hdl}
\end{equation}
\begin{equation}
	\mathbf{H}_{\mathrm{UL},k}=\mathbf{b}_\mathrm{BS}(-\vartheta_k)\mathbf{a}_\mathrm{RIS}(\phi_{\mathrm{r},k},\gamma_{\mathrm{r},k})^H,
	\label{equ_hul}
\end{equation}
where $\mathbf{a}_\mathrm{RIS}$, $\mathbf{a}_\mathrm{BS}$ and $\mathbf{b}_\mathrm{BS}$ are the steering vectors defined in what follows. We assume that the elements of each RIS are arranged in a uniform rectangular array whose size is $N_\mathrm{x}\times N_\mathrm{y}$, where $N_\mathrm{x}$ and $N_\mathrm{y}$ are the numbers of columns and rows, respectively. $\phi_{\mathrm{i},k}$ and $\gamma_{\mathrm{i},k}$ are the elevation and azimuth angles of arrival (AoA) and angles of departure (AoD) of the incident wave at the $k$-th RIS, respectively; $\phi_{\mathrm{r},k}$ and $\gamma_{\mathrm{r},k}$ are similarly defined for the reflected wave. The steering vector $\mathbf{a}_\mathrm{RIS}(\phi_{\mathrm{i/r},k},\gamma_{\mathrm{i/r},k})\in\mathbb{C}^{L\times 1}$ is defined as
\begin{equation}
	\mathbf{a}_\mathrm{RIS}(\phi_{\mathrm{i/r},k},\gamma_{\mathrm{i/r},k})=\mathbf{a}_\mathrm{x}(\phi_{\mathrm{i/r},k},\gamma_{\mathrm{i/r},k})\otimes\mathbf{a}_\mathrm{y}(\phi_{\mathrm{i/r},k},\gamma_{\mathrm{i/r},k}),
	\label{equ_asteer}
\end{equation}
where $\mathbf{a}_\mathrm{x}(\phi_{\mathrm{i/r},k},\gamma_{\mathrm{i/r},k})\in\mathbb{C}^{N_\mathrm{x}\times 1}$ and $\mathbf{a}_\mathrm{y}(\phi_{\mathrm{i/r},k},\gamma_{\mathrm{i/r},k})\in\mathbb{C}^{N_\mathrm{y}\times 1}$ are given as
	\begin{equation}\nonumber
		\mathbf{a}_\mathrm{x}(\phi_{\mathrm{i/r},k},\gamma_{\mathrm{i/r},k}) =\frac{1}{\sqrt{N_\mathrm{x}}} [1,\cdots,e^{j\pi (N_\mathrm{x}-1)\sin(\phi_{\mathrm{i/r},k})\cos(\gamma_{\mathrm{i/r},k})}]^T
	\end{equation}
	\begin{equation}\nonumber
		\mathbf{a}_\mathrm{y}(\phi_{\mathrm{i/r},k},\gamma_{\mathrm{i/r},k}) =\frac{1}{\sqrt{N_\mathrm{y}}} [1,\cdots,e^{j\pi (N_\mathrm{y}-1)\sin(\phi_{\mathrm{i/r},k})\sin(\gamma_{\mathrm{i/r},k})}]^T.
	\end{equation}
Moreover, we assume that both the transmit and receive antennas at the BS are arranged in a uniform linear antenna array. The steering vectors  $\mathbf{a}_\mathrm{BS}(\vartheta_k)\in\mathbb{C}^{N_\mathrm{t}\times 1}$ and $\mathbf{b}_\mathrm{BS}(-\vartheta_k)\in\mathbb{C}^{N_\mathrm{r}\times 1}$ are expressed as
\begin{equation}
	\mathbf{a}_\mathrm{BS}(\vartheta_k)=\frac{1}{\sqrt{N_\mathrm{t}}}[1,e^{j\pi\sin(\vartheta_k)},\cdots,e^{j\pi(N_\mathrm{t}-1)\sin(\vartheta_k)}]^T
\end{equation}
\begin{equation}
	\mathbf{b}_\mathrm{BS}(-\vartheta_k)=\frac{1}{\sqrt{N_\mathrm{r}}}[1,e^{j\pi\sin(-\vartheta_k)},\cdots,e^{j\pi(N_\mathrm{r}-1)\sin(-\vartheta_k)}]^T,
\end{equation}
where $\vartheta_k$ is the AoD to the $k$-th RIS from the BS. Since each RIS is designed to reflect the incident signal towards the opposite direction (i.e., the RIS operates as a retro-reflective metasurface for backscattering enhancement), the AoAs and the AoDs at the BS are opposite to each other. Defining 
\begin{equation}
	\alpha_{k,m} = \beta_k\mathbf{a}_\mathrm{RIS}(\phi_{\mathrm{r},k},\gamma_{\mathrm{r},k})^H\mathbf{\Lambda}_{k,m}\mathbf{a}_\mathrm{RIS}(\phi_{\mathrm{i},k},\gamma_{\mathrm{i},k}),
	\label{equ_alpha}
\end{equation}
we can rewrite \eqref{equ_ymt} as
\begin{equation}
	\mathbf{y}_m(t)=\sum_{k=0}^{K-1}\alpha_{k,m}\mathbf{H}(\vartheta_k)\mathbf{x}_m(t-\tau_k)+\mathbf{w}_m(t),
	\label{equ_sysmo}
\end{equation}
where $\mathbf{H}(\vartheta_k) = \mathbf{b}_\mathrm{BS}(-\vartheta_k)\mathbf{a}_\mathrm{BS}(\vartheta_k)^H\in\mathbb{C}^{N_\mathrm{r}\times N_\mathrm{t}}$.
For the $m$-th block, the demodulated signal at the $n$-th subcarrier is
\begin{equation}
	\mathbf{y}_m[n]=\frac{1}{T}\int_{mT_ \mathrm{b}+T_ \mathrm{cp}}^{(m+1)T_ \mathrm{b}}e^{-j2\pi n\Delta ft}\mathbf{y}_m(t)dt.
	\label{equ_fymt}
\end{equation}
Plugging \eqref{equ_xn} and \eqref{equ_sysmo} into \eqref{equ_fymt}, we have
\begin{equation}
	\mathbf{y}_m[n]
	=\sum_{k=0}^{K-1}\alpha_{k,m}e^{-j2\pi n\Delta f\tau_k}\mathbf{H}(\vartheta_k)\mathbf{x}_m[n]+\mathbf{w}_m[n],
	\label{equ_ysub}
\end{equation}
where 
\begin{equation}
	\nonumber
	\mathbf{w}_m[n]=\frac{1}{T}\int_{mT_\mathrm{b}+T_\mathrm{cp}}^{(m+1)T_\mathrm{b}}e^{-j2\pi n\Delta ft}\mathbf{w}_m(t)dt
\end{equation}
with $\mathbf{w}_m[n]\in\mathbb{C}^{N_\mathrm{r}\times 1}\sim\mathcal{CN}(\mathbf{0},\sigma^2\mathbf{I}_{N_\mathrm{r}})$.

\subsection{Passive Beamforming and Information Transfer of RIS}
\label{sec_passive}
\begin{figure}
	\centering
	\includegraphics[width=0.55\columnwidth]{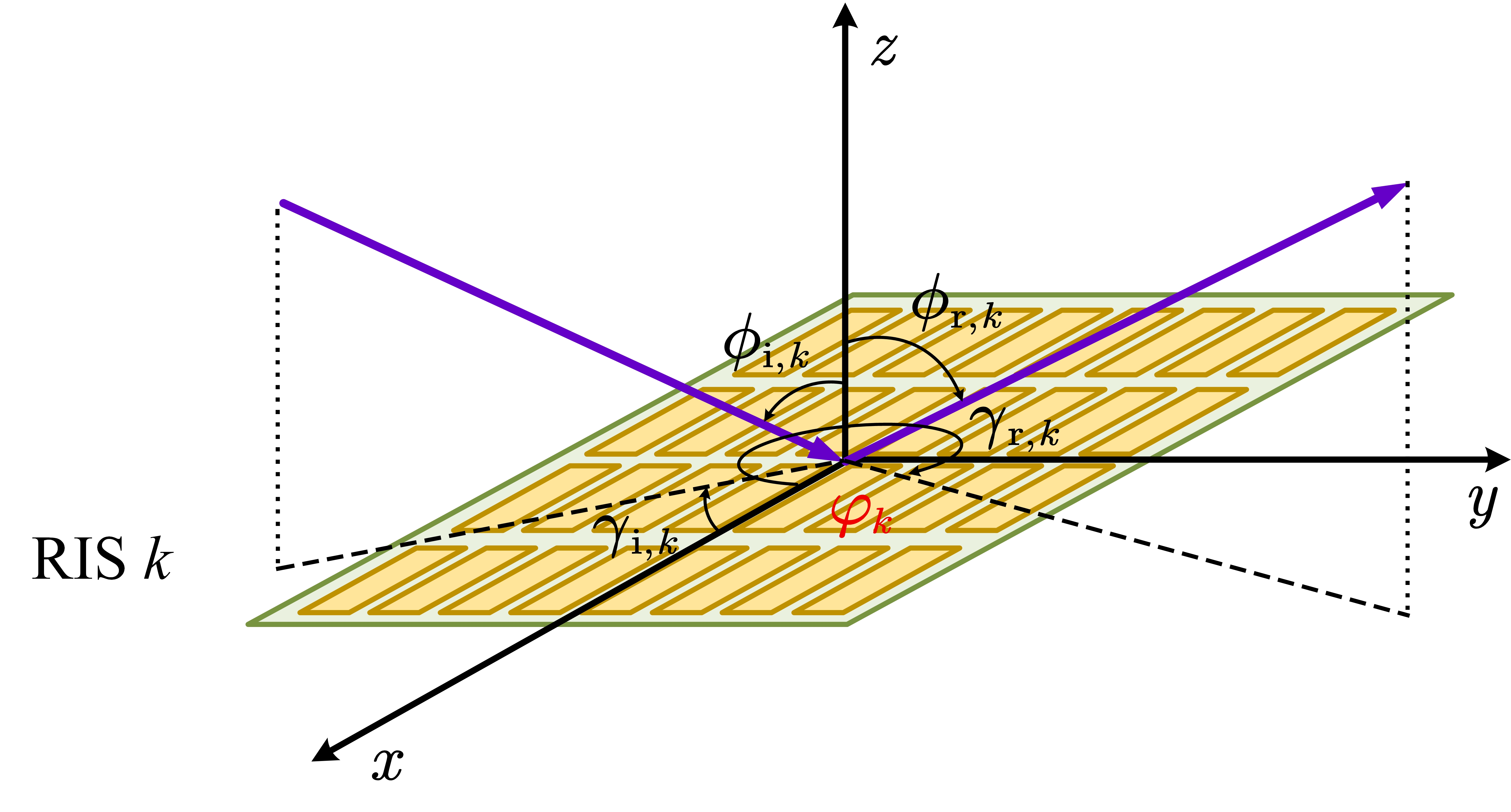}
	\caption{An illustration of the generalized Snell's law.}
	\label{fig_angle}
\end{figure}
We first illustrate an anomalous reflection scenario in Fig. \ref{fig_angle}, where $\phi_{\mathrm{i}}\in[-\frac{\pi}{2},\frac{\pi}{2}]$ and $\gamma_{\mathrm{i}}\in[0,2\pi]$ are the elevation and the azimuth angle of the incident wave, respectively; $\phi_{\mathrm{r}}\in[-\frac{\pi}{2},\frac{\pi}{2}]$ and $\gamma_{\mathrm{r}}\in[0,2\pi]$ are the elevation and the azimuth angle of the reflected wave, respectively. According to the generalized Snell’s law \cite{scien}, to reflect
an incident plane wave into a desired direction by breaking the specular reflection law (i.e., $\phi_{\mathrm{r}}= \phi_{\mathrm{i}}$, $\gamma_{\mathrm{r}}= \gamma_{\mathrm{i}}+\pi$),
the reflection phase of each element needs to be set linearly on the corresponding coordinates of both the $\mathrm{x}$- and $\mathrm{y}$-axes. Specifically, the phase shift of the ($i, j$)-th element $\theta_{k,m,i,j}$ of the $k$-th RIS is obtained by \cite{9445025}
\begin{equation}
	\theta_{k,m,i,j}=\pi (i-1)q_{\mathrm{x},k}+\pi (j-1)q_{\mathrm{y},k}+\varphi_{k,m}, 
	\label{eq_rislambda}
\end{equation}
where $\varphi_{k,m}$ is the reference phase at the origin of the coordinates; $q_{\mathrm{x},{k}}$ and $q_{\mathrm{y},{k}}$ denote the phase gradients of the
$\mathrm{x}$- and $\mathrm{y}$-axes, calculated as
\begin{equation}\label{equ_phasegra}
	\begin{bmatrix}
		q_{\mathrm{x},k}\\q_{\mathrm{y},k}
	\end{bmatrix}=\begin{bmatrix}
		\sin(\phi_{\mathrm{r},k})\cos(\gamma_{\mathrm{r},k})+\sin(\phi_{\mathrm{i},k})\cos(\gamma_{\mathrm{i},k})\\\sin(\phi_{\mathrm{r},k})\sin(\gamma_{\mathrm{r},k})+\sin(\phi_{\mathrm{i},k})\sin(\gamma_{\mathrm{i},k})
	\end{bmatrix}.
\end{equation}
Equation \eqref{equ_phasegra} reveals that the anomalous reflection from an arbitrary
$(\phi_{\mathrm{i},k},\gamma_{\mathrm{i},k})$ to an arbitrary $(\phi_{\mathrm{r},k},\gamma_{\mathrm{r},k})$ can be achieved by setting
$q_{\mathrm{x},k}$ and $q_{\mathrm{y},k}$ accordingly. Besides, another degree of freedom provided in \eqref{eq_rislambda} is the reference phase $\varphi_{k,m}$ of the $k$-th RIS, which determines the wavefront phase of the reflected beam in the $m$-th OFDM block. Particularly, for the scenario of retro-reflection (i.e., $\phi_{\mathrm{r},k}=-\phi_{\mathrm{i},k}$,  $\gamma_{\mathrm{r},k}=\gamma_{\mathrm{i},k}$), we have
\begin{equation}\label{equ_qxy}
	\begin{bmatrix}
		q_{\mathrm{x},k}\\q_{\mathrm{y},k}
	\end{bmatrix}=\begin{bmatrix}
		-2\sin(\phi_k)\cos(\gamma_k)\\-2\sin(\phi_k)\sin(\gamma_k)
	\end{bmatrix}.
\end{equation}
In practice, the angles $\phi_{\mathrm{i},k}$ and $\gamma_{\mathrm{i},k}$ can be estimated through sensors deployed at the $k$-th device, e.g., by using the MUSIC algorithm \cite{marco}. Plugging \eqref{eq_rislambda} and \eqref{equ_qxy} into \eqref{equ_alpha}, we have 
\begin{equation}
	\label{equ_betabeam}
		\begin{aligned}
			\alpha_{k,m} &=
			\frac{\beta_k}{L}\sum^{N_\mathrm{x}}_{i=1}\sum^{N_\mathrm{y}}_{l=1}e^{j2\pi(i-1)\sin(\phi_{\mathrm{r},k})\cos(\gamma_{\mathrm{r},k})}e^{j\theta_{k,m,i,l}} e^{j2\pi(l-1)\sin(\phi_{\mathrm{i},k})\sin(\gamma_{\mathrm{i},k})}\\&=\beta_ke^{j\varphi_{k,m}},
		\end{aligned}
\end{equation}
where $\varphi_{k,m}$ can be easily controlled by the $k$-th RIS. This motivates us to apply the reflection modulation on the incident wave \cite{marco352}. The reflected wave, or more specifically, $\varphi_{k,m}$, can be utilized to passively transmit new information generated at device to the BS. 

Since the localization parameters are assumed to be constant in $M$ consecutive OFDM blocks, the complex path coefficient $\beta_k$ is a shared constant. Then, \eqref{equ_betabeam} is extended to 
\begin{equation}\label{eq_shared}
    [\alpha_{k,1},\cdots,\alpha_{k,M}]=\beta_k[e^{j\varphi_{k,1}},\cdots,e^{j\varphi_{k,M}}].
\end{equation}
To retrieve information from the phase of $\alpha_{k,m}$, i.e. $\angle\alpha_{k,m}=\angle\beta_k+\varphi_{k,m}$, the differential phase shift keying (DPSK) modulation is used to cancel out the unknown common phase $\angle\beta_k$ caused by the environment during the signal transmission. The modulation process is described as
\begin{equation}\label{equ_modu}
	\varphi_{k,m} = \varphi_{k,m-1} + S_{k,m} + S_\text{ref},\qquad m\in[2,M]
\end{equation}
where $S_{k,m}\in\{S_1,\cdots,S_V\}$ is the modulated phase difference with $\{S_1,\cdots,S_V\}$ being the set of $V$ possible phase differences in DPSK, and $S_\text{ref}$ is the known common reference phase to ensure $\varphi_{k,m}\in\{S_1,\cdots,S_V\}$. For example, when differential quadrature phase shift keying (DQPSK) is applied, $S_{k,m}\in\{\frac{\pi}{4},\frac{3\pi}{4},\frac{5\pi}{4},\frac{7\pi}{4}\}$ and $S_\text{ref}=\frac{\pi}{4}$.

Recall that $\{\mathbf{x}_m[n]\}$ is the self-information known by the BS. Then, the problem interested here is to jointly estimate the unknown parameters $\vartheta_k$,  $\tau_k$ and $\alpha_{k,m}$ from the noisy observations $\{\mathbf{y}_m[n]\}$ in \eqref{equ_ysub}. After obtaining the estimated $\hat{\alpha}_{k,m}$, the estimated phase difference $\hat{S}_{k,m}$ is calculated as
\begin{equation}
	\hat{S}_{k,m} = \angle\hat{\alpha}_{k,m}-\angle\hat{\alpha}_{k,m-1}- S_\text{ref},\qquad m\in[2,M]
\end{equation}
where $\angle\beta_k$ is canceled during the subtraction.

\section{Joint Localization and Information Recovery}\label{sec3}
In this section, we first introduce the grid-based parametric model to formulate the problem stated in the preceding section, and then present a novel message-passing algorithm to solve the problem.

\subsection{Parametric Grid Model}
The service ranges of angles and delays are first quantized to $Q$ and $U$ discrete values as
\begin{subequations}
	\label{equ_grids}
	\begin{equation}
		\centering
		\bm{\vartheta} = [\vartheta_0,\cdots,\vartheta_{Q-1}]^T,\hspace{1.3em} \vartheta_q\in[\vartheta_{\min}, \vartheta_{\max}]
	\end{equation}
	\begin{equation}
		\centering
		\bm{\tau} = [\tau_0,\cdots,\tau_{U-1}]^T,\hspace{1.5em} \tau_u\in[\tau_{\min}, \tau_{\max}],
	\end{equation}
\end{subequations}
where $\vartheta_0 = \vartheta_{\min}$, $\vartheta_{Q-1} = \vartheta_{\max}$, $\tau_0 = \tau_{\min}$, $\tau_{U-1} = \tau_{\max}$. A uniform grid is employed in this work, and the differences between adjacent angles and delays are $\frac{\vartheta_{\max}-\vartheta_{\min}}{Q}$ and $\frac{\tau_{\max}-\tau_{\min}}{U}$, respectively, with $Q\gg K$ and $U\gg K$ to achieve the desired resolution. Based on the above discrete angles and delays, \eqref{equ_ysub} is rewritten as 
\begin{equation}\label{equ_sysmodmn}
	\begin{aligned}
		\mathbf{y}_m[n]&=\sum_{u=0}^{U-1}\sum_{q=0}^{Q-1}e^{-j2\pi n\Delta f\tau_u}\mathbf{H}(\vartheta_q)\mathbf{x}_m[n]\zeta_{q,u,m}+\mathbf{w}_m[n]\\&=\mathbf{Z}_m[n]\bm{\zeta}_m+\mathbf{w}_m[n]
	\end{aligned}
\end{equation}
with 
\begin{equation}
	\mathbf{Z}_m[n] = (e^{-j2\pi n\Delta f\bm{\tau}})^T \otimes \mathbf{H}(\bm{\vartheta})(\mathbf{I}_Q\otimes\mathbf{x}_m[n]),
	\label{equ_Z}
\end{equation}
where 
$e^{-j2\pi n\Delta f\bm{\tau}}=[e^{-j2\pi n\Delta f\tau_0},\cdots,e^{-j2\pi n\Delta f\tau_{U-1}}]^T\in\mathbb{C}^{U\times 1}$ and  $\mathbf{H}(\bm{\vartheta})=[\mathbf{H}(\vartheta_0),\cdots,\mathbf{H}(\vartheta_{Q-1})]\in\mathbb{C}^{N_\mathrm{r}\times QN_\mathrm{t}}$. $\bm{\zeta}_m=[\zeta_{1,1,m},\cdots,\zeta_{Q,1,m},\zeta_{1,2,m},\cdots,\zeta_{Q,U,m}]^T\in\mathbb{C}^{QU\times1}$, where $\zeta_{q,u,m}$ corresponds to a possible device with the angle $\vartheta_q$ and the delay $\tau_u$ in the $m$-th OFDM block. Since there are only $K$ devices in the considered region, $\bm{\zeta}_m$ is a sparse vector, where the non-zero elements are given by $\{\alpha_{k,m}|k\in[1,K]\}$.

Collecting the received signals on $N$ subcarriers, we obtain 
\begin{equation}
	\begin{aligned}
		\mathbf{y}_m =[\mathbf{y}_m[1]^T,\cdots,\mathbf{y}_m[N]^T]^T= \mathbf{Z}_m\bm{\zeta}_m + \mathbf{w}_m,
	\end{aligned}
\end{equation}
where $\mathbf{Z}_m=[\mathbf{Z}_m[1]^T,\cdots,\mathbf{Z}_m[N]^T]^T\in\mathbb{C}^{N_\mathrm{r}N\times UQ}$ and $\mathbf{w}_m=[\mathbf{w}_m[1]^T,\cdots,\mathbf{w}_m[N]^T]^T\in\mathbb{C}^{N_\mathrm{r}N\times1}$. Considering all the demodulated symbols in one transmission frame,
we have 
\begin{equation}
	\mathbf{Y}=[\mathbf{y}_1,\cdots,\mathbf{y}_M]= [\mathbf{Z}_1\bm{\zeta}_1,\cdots,\mathbf{Z}_M\bm{\zeta}_M] + \mathbf{W},
	\label{equ_sysmode}
\end{equation}
where $\mathbf{W}=[\mathbf{w}_1,\cdots,\mathbf{w}_M]\in\mathbb{C}^{N_\mathrm{r}N\times M}$. 

With the knowledge of the transmitted symbols $\mathbf{x}_m[n]$ and the grids in \eqref{equ_grids}, the matrix $\mathbf{Z}_m$ is known by the BS. The estimation problem is converted into the problem of estimating $\bm{\zeta}=[\bm{\zeta}_1,\cdots,\bm{\zeta}_M]\in\mathbb{C}^{UQ\times M}$ based on $\mathbf{Y}$ in \eqref{equ_sysmode}. 

\subsection{Message-Passing Algorithm}
The minimum mean square error (MMSE) estimator of $\bm{\zeta}$ given $\mathbf{Y}$ is given by
\begin{equation}
	\hat{\bm{\zeta}} = E\{\bm{\zeta}|\mathbf{Y}\}.
	\label{equ_MMSE}
\end{equation}
However, the calculation of the conditional mean $E\{\bm{\zeta}|\mathbf{Y}\}$ is computationally involving. In the following, we develop a computationally efficient iterative algorithm to approximately calculate the MMSE estimator \eqref{equ_MMSE} based on the message-passing rule. 

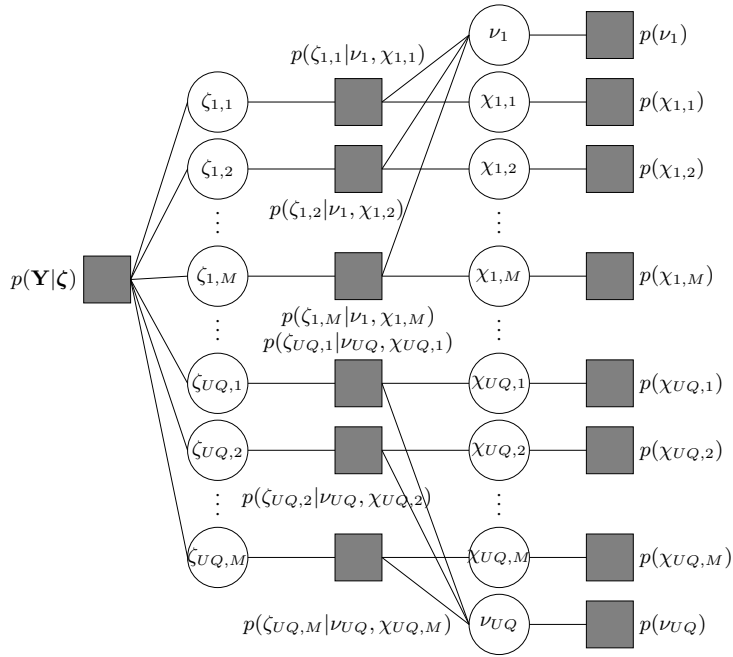
\begin{figure}[!htbp]
	\centering
	\resizebox{0.55\textwidth}{!}{\begin{tikzpicture}
	\draw[fill=gray] (0,0) rectangle node[left=0.3cm] {$p(\mathbf{Y}|\bm{\zeta})$} (0.7,0.7);
	\draw (0.7,0.35) -- (2-0.45,3);
	\draw (0.7,0.35) -- (2-0.45,2);
	\draw (0.7,0.35) -- (2-0.45,0.4);
	\draw (0.7,0.35) -- (2-0.45,-1.2);
	\draw (0.7,0.35) -- (2-0.45,-2.2);
	\draw (0.7,0.35) -- (2-0.45,-3.8);
	
	\draw (2,3) node (l1) {$\zeta_{1,1}$} circle [radius=0.45];
	\draw (2,2) node (l2) {$\zeta_{1,2}$} circle [radius=0.45];
	\draw (2,1.3) node (l3) {$\vdots$};
	\draw (2,0.4) node (l4) {$\zeta_{1,M}$} circle [radius=0.45];
	\draw (2,-0.3) node (l9) {$\vdots$};
	\draw (2,-1.2) node (l5) {$\zeta_{UQ,1}$} circle [radius=0.45];
	\draw (2,-2.2) node (l6) {$\zeta_{UQ,2}$} circle [radius=0.45];
	\draw (2,-2.9) node (l7) {$\vdots$};
	\draw (2,-3.8) node (l8) {$\zeta_{UQ,M}$} circle [radius=0.45];
	
	\draw (2.45,3) -- (3.75,3);
	\draw (2.45,2) -- (3.75,2);
	\draw (2.45,0.4) -- (3.75,0.4);
	\draw (2.45,-1.2) -- (3.75,-1.2);
	\draw (2.45,-2.2) -- (3.75,-2.2);
	\draw (2.45,-3.8) -- (3.75,-3.8);
	
	\draw[fill=gray] (3.75,2.65) rectangle node[above=0.4cm] {$p(\zeta_{1,1}|\nu_1,\chi_{1,1})$} (4.45,3.35);
	\draw[fill=gray] (3.75,1.65) rectangle node[below=0.3cm] {$p(\zeta_{1,2}|\nu_1,\chi_{1,2})\qquad$} (4.45,2.35);
	\draw[fill=gray] (3.75,0.05) rectangle node[below=0.3cm] {$p(\zeta_{1,M}|\nu_1,\chi_{1,M})$} (4.45,0.75);
	\draw[fill=gray] (3.75,-1.55) rectangle node[above=0.3cm] {$p(\zeta_{UQ,1}|\nu_{UQ},\chi_{UQ,1})$} (4.45,-0.85);
	\draw[fill=gray] (3.75,-2.55) rectangle node[below=0.4cm] {$p(\zeta_{UQ,2}|\nu_{UQ},\chi_{UQ,2})\qquad$} (4.45,-1.85);
	\draw[fill=gray] (3.75,-4.15) rectangle node[below=0.7cm] {$p(\zeta_{UQ,M}|\nu_{UQ},\chi_{UQ,M})\quad$} (4.45,-3.45);
	
	\draw (4.45,3) -- (5.75,3);
	\draw (4.45,2) -- (5.75,2);
	\draw (4.45,0.4) -- (5.75,0.4);
	\draw (4.45,-1.2) -- (5.75,-1.2);
	\draw (4.45,-2.2) -- (5.75,-2.2);
	\draw (4.45,-3.8) -- (5.75,-3.8);
	
	\draw (6.2,3) node (l1) {$\chi_{1,1}$} circle [radius=0.45];
	\draw (6.2,2) node (l2) {$\chi_{1,2}$} circle [radius=0.45];
	\draw (6.2,1.3) node (l3) {$\vdots$};
	\draw (6.2,0.4) node (l4) {$\chi_{1,M}$} circle [radius=0.45];
	\draw (6.2,-0.3) node (l9) {$\vdots$};
	\draw (6.2,-1.2) node (l5) {$\chi_{UQ,1}$} circle [radius=0.45];
	\draw (6.2,-2.2) node (l6) {$\chi_{UQ,2}$} circle [radius=0.45];
	\draw (6.2,-2.9) node (l7) {$\vdots$};
	\draw (6.2,-3.8) node (l8) {$\chi_{UQ,M}$} circle [radius=0.45];
	
	\draw (6.65,3) -- (7.6,3);
	\draw (6.65,2) -- (7.6,2);
	\draw (6.65,0.4) -- (7.6,0.4);
	\draw (6.65,-1.2) -- (7.6,-1.2);
	\draw (6.65,-2.2) -- (7.6,-2.2);
	\draw (6.65,-3.8) -- (7.6,-3.8);
	
	\draw[fill=gray] (7.5,2.65) rectangle node[right=0.3cm] {$p(\chi_{1,1})$} (8.2,3.35);
	\draw[fill=gray] (7.5,1.65) rectangle node[right=0.3cm] {$p(\chi_{1,2})$} (8.2,2.35);
	\draw[fill=gray] (7.5,0.05) rectangle node[right=0.3cm] {$p(\chi_{1,M})$} (8.2,0.75);
	\draw[fill=gray] (7.5,-1.55) rectangle node[right=0.3cm] {$p(\chi_{UQ,1})$} (8.2,-0.85);
	\draw[fill=gray] (7.5,-2.55) rectangle node[right=0.3cm] {$p(\chi_{UQ,2})$} (8.2,-1.85);
	\draw[fill=gray] (7.5,-4.15) rectangle node[right=0.3cm] {$p(\chi_{UQ,M})$} (8.2,-3.45);
	
	\draw (6.2,4) node (l1) {$\nu_{1}$} circle [radius=0.45];
	\draw[fill=gray] (7.5,3.65) rectangle node[right=0.3cm] {$p(\nu_{1})$} (8.2,4.35);
	\draw (6.65,4) -- (7.5,4);
	
	\draw (4.45,3) -- (5.75,4);
	\draw (4.45,2) -- (5.75,4);
	\draw (4.45,0.4) -- (5.75,4);

	\draw (6.2,-4.8) node (l1) {$\nu_{UQ}$} circle [radius=0.45];
	\draw[fill=gray] (7.5,-5.15) rectangle node[right=0.3cm] {$p(\nu_{UQ})$} (8.2,-4.45);
	\draw (6.65,-4.8) -- (7.5,-4.8);

	\draw (4.45,-1.2) -- (5.75,-4.8);
	\draw (4.45,-2.2) -- (5.75,-4.8);
	\draw (4.45,-3.8) -- (5.75,-4.8);
	
\end{tikzpicture}}
	\caption{The factor graph of the proposed algorithm.}
	\label{fig_factor}
\end{figure}

From \eqref{eq_shared}, each non-zero row of $\bm{\zeta}$ is the multiplication of a shared constant $\beta_k$ and phase shifts in different OFDM blocks $[e^{j\varphi_{k,1}},\cdots,e^{j\varphi_{k,M}}]$. Then we define $\bm{\zeta} = \text{diag}(\bm{\nu})\bm{\chi}$, where $\beta_k$ and $[e^{j\varphi_{k,1}},\cdots,e^{j\varphi_{k,M}}]$ can be found as the non-zero elements in $\bm{\nu}\in\mathbb{C}^{UQ\times 1}$ and non-zero rows in $\bm{\chi}\in\mathbb{C}^{UQ\times M}$, respectively.
The posterior probability $p(\bm{\zeta}|\mathbf{Y})$ is factorized as
\begin{equation}
		p(\bm{\zeta}|\mathbf{Y})=\frac{1}{p(\mathbf{Y})}p(\mathbf{Y}|\bm{\zeta})p(\bm{\zeta}|\bm{\nu},\bm{\chi})p(\bm{\nu})p(\bm{\chi}),
	\label{equ_post}
\end{equation}
where
\begin{align}
	p(\mathbf{Y}|\bm{\zeta})&=\prod_{m=1}^{M}\mathcal{CN}(\mathbf{y}_m;\mathbf{Z}_m\bm{\zeta}_m,\sigma^2\mathbf{I}_{N_\mathrm{r}N})\label{equ_yzeta}\\
	p(\bm{\zeta}|\bm{\nu},\bm{\chi}) &= \delta(\bm{\zeta}-\text{diag}(\bm{\nu})\bm{\chi})
	\label{equ_beta} \\
	p(\bm{\nu}) &= \prod_{i=1}^{UQ} \left((1-\rho)\delta(\nu_i)+\rho\mathcal{CN}(\nu_i;\hat{\beta},v^{\beta})\right)
	\label{equ_nu} \\
	p(\bm{\chi}) &= \prod_{i=1}^{UQ}\prod_{m=1}^{M} \left(\sum_{l=1}^{V}\frac{1}{V}\delta(\chi_{i,m}-e^{jS_l})\right).
	\label{equ_chi}
\end{align} 
In the above, $\delta(\cdot)$ is the Dirac delta function; $\rho=\frac{K}{UQ}$ is the sparsity of $\bm{\nu}$ (or the row sparsity of $\bm{\zeta}$); $p(\beta_k)$ is approximated as Gaussian distribution with mean $\hat{\beta}$ and variance $v^{\beta}$; $\chi_{i,m}$ is the element of $\bm{\chi}$ in the $i$-th row and the $m$-th column; $S_l$ denotes the $l$-th phase shift within the set of $V$ possible phases in DPSK, where each phase has the probability of $\frac{1}{V}$. The factor graph representation of $p(\bm{\zeta}|\mathbf{Y})$ is shown in Fig. \ref{fig_factor}. The hollow circles and the solid squares represent the variable nodes and the factor nodes, respectively. Detailed message passings are described as follows.

\subsubsection{Forward Message Passing}
Given the message $\mathcal{M}_{\zeta_{i,m}\rightarrow p(\zeta_{i,m}|\nu_i,\chi_{i,m})}(\zeta_{i,m})\sim \mathcal{CN}(\zeta_{i,m};\hat{r}_{i,m},v^r_{i,m})$, where $\hat{r}_{i,m}$ and $v^r_{i,m}$ are the outputs of the input linear step in the generalized approximate message passing (GAMP) algorithm \cite{6556987} (more details can be found at the end of this subsection),  the message from factor node $p(\zeta_{i,m}|\nu_i,\chi_{i,m})$ to variable node $\nu_i$ is calculated from the sum-product rule as
\begin{equation}
	\mathcal{M}_{p(\zeta_{i,m}|\nu_i,\chi_{i,m})\rightarrow\nu_i}(\nu_i) = \int_{\zeta_{i,m}}\sum_{\chi_{i,m}}p(\zeta_{i,m}|\nu_i,\chi_{i,m})\mathcal{M}_{\zeta_{i,m}\rightarrow p(\zeta_{i,m}|\nu_i,\chi_{i,m})}(\zeta_{i,m})\mathcal{M}_{\chi_{i,m}\rightarrow p(\zeta_{i,m}|\nu_i,\chi_{i,m})}(\chi_{i,m}).
	\label{equ_m1}
\end{equation}
Inserting \eqref{equ_beta} and the prior probability \eqref{equ_chi} into \eqref{equ_m1}, we have
\begin{equation}\label{equ_up}
	\begin{aligned}
		\mathcal{M}_{p(\zeta_{i,m}|\nu_i,\chi_{i,m})\rightarrow\nu_i}(\nu_i) &= \int_{\zeta_{i,m}}\sum_{\chi_{i,m}}\delta(\zeta_{i,m}-\nu_i\chi_{i,m})\mathcal{CN}(\zeta_{i,m};\hat{r}_{i,m},v^r_{i,m})\sum_{l=1}^{V}\frac{1}{V}\delta(\chi_{i,m}-e^{jS_l})
		\\&=\sum_{l=1}^{V}\frac{1}{V}\mathcal{CN}(\nu_{i};e^{-jS_l}\hat{r}_{i,m},v^r_{i,m}).
	\end{aligned}
\end{equation}

\subsubsection{Backward Message Passing}
With the Bernoulli Gaussian prior \eqref{equ_nu} and the Gaussian mixture \eqref{equ_up}, the message from variable node $\nu_i$ to factor node $p(\zeta_{i,m}|\nu_i,\chi_{i,m})$ is calculated as 
\begin{equation}\label{equ_mvp}
	\begin{aligned}
		\mathcal{M}_{\nu_i\rightarrow p(\zeta_{i,m}|\nu_i,\chi_{i,m})}(\nu_i)&=\mathcal{M}_{p(\nu_i)\rightarrow\nu_i}(\nu_i)\prod_{m'=1,m'\neq m}^{M}\mathcal{M}_{p(\zeta_{i,m'}|\nu_i,\chi_{i,m'})\rightarrow\nu_i}(\nu_i)
		\\&=\left((1-\rho)\delta(\nu_i)+\rho\mathcal{CN}(\nu_i;\hat{\beta},v^{\beta})\right)\prod_{m'=1,m'\neq m}^{M}\left(\sum_{l=1}^{V}\frac{1}{V}\mathcal{CN}(\nu_{i};e^{-jS_l}\hat{r}_{i,m'},v^r_{i,m'})\right).
	\end{aligned}
\end{equation}
Note that the exact form of \eqref{equ_mvp} is difficult to calculate due to the product of multiple Gaussian mixtures. The number of overall Gaussian components grows exponentially with the number of OFDM blocks $M$. To alleviate this problem, we propose a progressive approximation method to simplify the product of the Gaussian mixtures as follows.

The multiplication between the $m'$-th and the $(m'+1)$-th Gaussian mixture in \eqref{equ_mvp} is calculated as
\begin{equation}\label{equ_exactmultiplication}
	\begin{aligned}
		&\left(\sum_{l=1}^{V}\frac{1}{V}\mathcal{CN}(\nu_{i};e^{-jS_l}\hat{r}_{i,m'},v^r_{i,m'})\right)\left(\sum_{q=1}^{V}\frac{1}{V}\mathcal{CN}(\nu_{i};e^{-jS_q}\hat{r}_{i,m'+1},v^r_{i,m'+1})\right)\\=&\sum_{l=1}^{V}\sum_{q=1}^{V}\xi_{l,q}\mathcal{CN}(\nu_{i};\hat{\nu}_{i,m'+1,l,q},v^{\nu}_{i,m'+1})
	\end{aligned}
\end{equation}
where from the Gaussian message combining property\footnote{The multiplication of two Gaussian functions is another Gaussian function: $\mathcal{CN}(x;a,A)\mathcal{CN}(x;b,B)=d\mathcal{CN}(x;c,C)$ with $d=\mathcal{CN}(0;a-b,A+B)$, $C=(A^{-1}+B^{-1})^{-1}$ and $c=C(a/A+b/B)$.} the variance $v^{\nu}_{i,m'+1}$, the mean $\hat{\nu}_{i,m'+1,l,q}$ and the weight $\xi_{l,q}$ are respectively given by
\begin{subequations}\label{equ_weight}
	\begin{align}
		v^{\nu}_{i,m'+1} &= \left(\frac{1}{v^{r}_{i,m'}}+\frac{1}{v^{r}_{i,m'+1}}\right)^{-1}\label{equ_v1}\\
		\hat{\nu}_{i,m'+1,l,q} &= v^{\nu}_{i,m'+1}\left(e^{-jS_l}\frac{\hat{r}_{i,m'}}{v^{r}_{i,m'}}+e^{-jS_q}\frac{\hat{r}_{i,m'+1}}{v^{r}_{i,m'+1}}\right)\\
		\xi_{l,q}&=\frac{\mathcal{CN}(0;e^{-jS_l}\hat{r}_{i,m'}-e^{-jS_q}\hat{r}_{i,m'+1},v^{r}_{i,m'}+v^{r}_{i,m'+1})}{\sum_{l'=1}^{V}\sum_{q'=1}^{V}\mathcal{CN}(0;e^{-jS_{l'}}\hat{r}_{i,m'}-e^{-jS_{q'}}\hat{r}_{i,m'+1},v^{r}_{i,m'}+v^{r}_{i,m'+1})}.
	\end{align}
\end{subequations}
Since each Gaussian component in \eqref{equ_exactmultiplication} shares the same variances $v^{\nu}_{i,m'+1}$, we cluster the mean values into $V$ groups and approximate the values of each group as one mean value. Clustering algorithms, such as K-means \cite{kmeans} and expectation maximization (EM) \cite{em}, can be used for this purpose.
In this work, we group the mean values $\{\hat{\nu}_{i,m'+1,l,q}\}$ as follows.

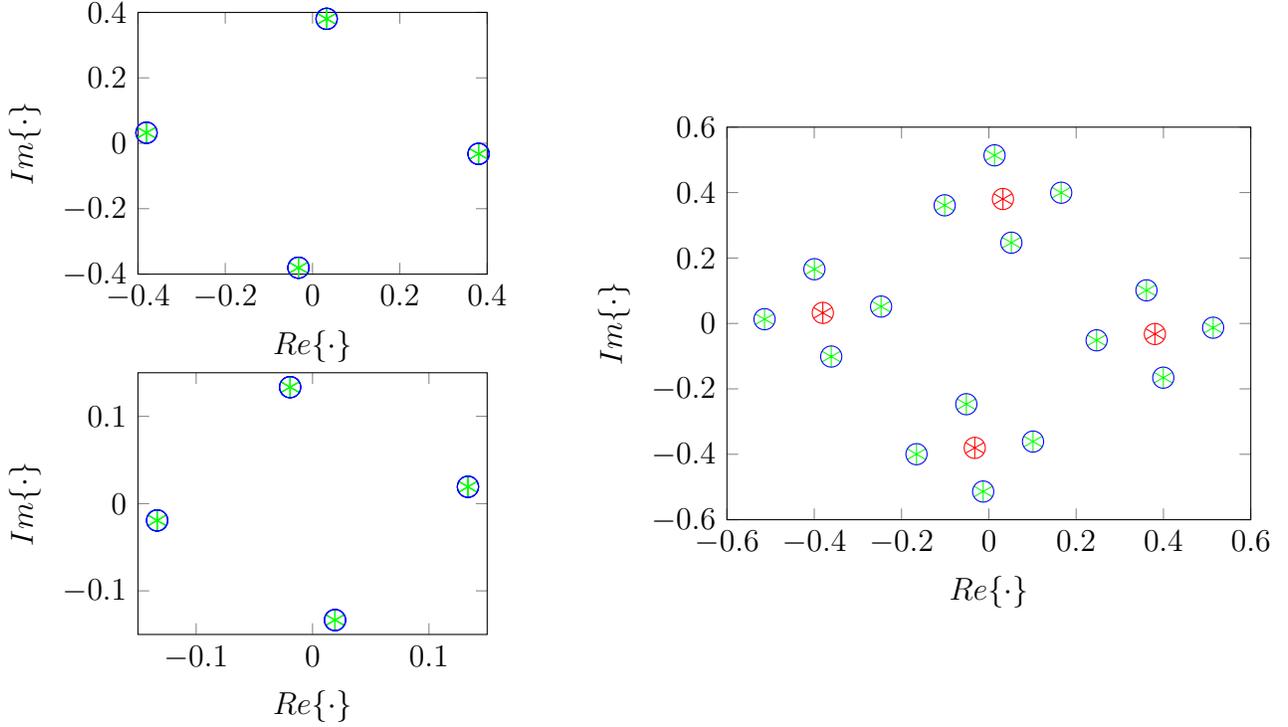
\begin{figure}[!htbp]
	\begin{minipage}{0.32\columnwidth}
		\hspace{0.3cm}\begin{subfigure}{\columnwidth}
			\pgfplotsset{every axis label/.append style={font=\normalsize},
	every tick label/.append style={font=\normalsize},
}

\begin{tikzpicture}
	
	\begin{axis}[%
		width=.8\columnwidth,
		height=.6\columnwidth,
		at={(0.758in,0.603in)},
		scale only axis,
		xlabel style={font=\normalsize},
		ylabel style={font=\normalsize},
		xlabel={$Re\{\cdot\}$},
		ylabel={$Im\{\cdot\}$},
		xmin=-0.4,
		xmax=0.4,
		ymin=-0.4,
		ymax=0.4,
		axis background/.style={fill=white},
		]
		\addplot [color=green, only marks, mark=asterisk, mark options={solid, green}, mark size=4pt]
		table[row sep=crcr]{%
			0.380356333680647	-0.0322951772183134\\
			0.0322951772183134	0.380356333680647\\
			-0.380356333680647	0.0322951772183134\\
			-0.0322951772183134	-0.380356333680647\\
			0.380356333680647	-0.0322951772183134\\
			0.0322951772183134	0.380356333680647\\
			-0.380356333680647	0.0322951772183134\\
			-0.0322951772183134	-0.380356333680647\\
			0.380356333680647	-0.0322951772183134\\
			0.0322951772183134	0.380356333680647\\
			-0.380356333680647	0.0322951772183134\\
			-0.0322951772183134	-0.380356333680647\\
			0.380356333680647	-0.0322951772183134\\
			0.0322951772183134	0.380356333680647\\
			-0.380356333680647	0.0322951772183134\\
			-0.0322951772183134	-0.380356333680647\\
		};
		\addplot [color=blue, only marks, mark=o, mark options={solid, blue}, mark size=4pt]
		table[row sep=crcr]{%
			0.380356333680647	-0.0322951772183134\\
			0.0322951772183134	0.380356333680647\\
			-0.380356333680647	0.0322951772183134\\
			-0.0322951772183134	-0.380356333680647\\
			0.380356333680647	-0.0322951772183134\\
			0.0322951772183134	0.380356333680647\\
			-0.380356333680647	0.0322951772183134\\
			-0.0322951772183134	-0.380356333680647\\
			0.380356333680647	-0.0322951772183134\\
			0.0322951772183134	0.380356333680647\\
			-0.380356333680647	0.0322951772183134\\
			-0.0322951772183134	-0.380356333680647\\
			0.380356333680647	-0.0322951772183134\\
			0.0322951772183134	0.380356333680647\\
			-0.380356333680647	0.0322951772183134\\
			-0.0322951772183134	-0.380356333680647\\
		};
	\end{axis}
\end{tikzpicture}%
		\end{subfigure}\vfill
		\hspace{0.3cm}\begin{subfigure}{\columnwidth}
			\pgfplotsset{every axis label/.append style={font=\normalsize},
	every tick label/.append style={font=\normalsize},
}

\begin{tikzpicture}
	
	\begin{axis}[%
		width=.8\columnwidth,
		height=.6\columnwidth,
		at={(0.758in,0.603in)},
		scale only axis,
		xlabel style={font=\normalsize},
		ylabel style={font=\normalsize},
		xlabel={$Re\{\cdot\}$},
		ylabel={$Im\{\cdot\}$},
		xmin=-0.15,
		xmax=0.15,
		ymin=-0.15,
		ymax=0.15,
		axis background/.style={fill=white},
		]
		\addplot [color=green, only marks, mark=asterisk, mark options={solid, green}, mark size=4pt]
		table[row sep=crcr]{%
			-0.0192860315098576	0.133489279813782\\
			-0.0192860315098576	0.133489279813782\\
			-0.0192860315098576	0.133489279813782\\
			-0.0192860315098576	0.133489279813782\\
			0.133489279813782	0.0192860315098576\\
			0.133489279813782	0.0192860315098576\\
			0.133489279813782	0.0192860315098576\\
			0.133489279813782	0.0192860315098576\\
			0.0192860315098576	-0.133489279813782\\
			0.0192860315098576	-0.133489279813782\\
			0.0192860315098576	-0.133489279813782\\
			0.0192860315098576	-0.133489279813782\\
			-0.133489279813782	-0.0192860315098576\\
			-0.133489279813782	-0.0192860315098576\\
			-0.133489279813782	-0.0192860315098576\\
			-0.133489279813782	-0.0192860315098576\\
		};
		\addplot [color=blue, only marks, mark=o, mark options={solid, blue}, mark size=4pt]
		table[row sep=crcr]{%
			-0.0192860315098576	0.133489279813782\\
			-0.0192860315098576	0.133489279813782\\
			-0.0192860315098576	0.133489279813782\\
			-0.0192860315098576	0.133489279813782\\
			0.133489279813782	0.0192860315098576\\
			0.133489279813782	0.0192860315098576\\
			0.133489279813782	0.0192860315098576\\
			0.133489279813782	0.0192860315098576\\
			0.0192860315098576	-0.133489279813782\\
			0.0192860315098576	-0.133489279813782\\
			0.0192860315098576	-0.133489279813782\\
			0.0192860315098576	-0.133489279813782\\
			-0.133489279813782	-0.0192860315098576\\
			-0.133489279813782	-0.0192860315098576\\
			-0.133489279813782	-0.0192860315098576\\
			-0.133489279813782	-0.0192860315098576\\
		};
	\end{axis}
\end{tikzpicture}%
		\end{subfigure}
	\end{minipage}\hfill
	\begin{minipage}{0.48\columnwidth}
		\hspace{-1.3cm}\begin{subfigure}{\columnwidth}
			\pgfplotsset{every axis label/.append style={font=\normalsize},
	every tick label/.append style={font=\normalsize},
}

\begin{tikzpicture}
	
	\begin{axis}[%
		width=.8\columnwidth,
		height=.6\columnwidth,
		at={(0.758in,0.603in)},
		scale only axis,
		xlabel style={font=\normalsize},
		ylabel style={font=\normalsize},
		xlabel={$Re\{\cdot\}$},
		ylabel={$Im\{\cdot\}$},
		xmin=-0.6,
		xmax=0.6,
		ymin=-0.6,
		ymax=0.6,
		axis background/.style={fill=white},
		]
		\addplot [color=green, only marks, mark=asterisk, mark options={solid, green}, mark size=4pt]
		table[row sep=crcr]{%
			0.36107030217079	0.101194102595469\\
			0.0130091457084558	0.51384561349443\\
			-0.399642365190505	0.165784457032096\\
			-0.051581208728171	-0.246867053866865\\
			0.51384561349443	-0.0130091457084558\\
			0.165784457032096	0.399642365190505\\
			-0.246867053866865	0.051581208728171\\
			0.101194102595469	-0.36107030217079\\
			0.399642365190505	-0.165784457032096\\
			0.051581208728171	0.246867053866865\\
			-0.36107030217079	-0.101194102595469\\
			-0.0130091457084558	-0.51384561349443\\
			0.246867053866865	-0.051581208728171\\
			-0.101194102595469	0.36107030217079\\
			-0.51384561349443	0.0130091457084558\\
			-0.165784457032096	-0.399642365190505\\
		};
		\addplot [color=blue, only marks, mark=o, mark options={solid, blue}, mark size=4pt]
		table[row sep=crcr]{%
			0.36107030217079	0.101194102595469\\
			0.0130091457084558	0.51384561349443\\
			-0.399642365190505	0.165784457032096\\
			-0.051581208728171	-0.246867053866865\\
			0.51384561349443	-0.0130091457084558\\
			0.165784457032096	0.399642365190505\\
			-0.246867053866865	0.051581208728171\\
			0.101194102595469	-0.36107030217079\\
			0.399642365190505	-0.165784457032096\\
			0.051581208728171	0.246867053866865\\
			-0.36107030217079	-0.101194102595469\\
			-0.0130091457084558	-0.51384561349443\\
			0.246867053866865	-0.051581208728171\\
			-0.101194102595469	0.36107030217079\\
			-0.51384561349443	0.0130091457084558\\
			-0.165784457032096	-0.399642365190505\\
		};
		\addplot [color=red, only marks, mark=asterisk, mark options={solid, red}, mark size=4pt]
		table[row sep=crcr]{%
			0.38036099394506	-0.0323020684174378\\
			0.0323020684174378	0.38036099394506\\
			-0.38036099394506	0.0323020684174378\\
			-0.0323020684174378	-0.38036099394506\\
		};
		\addplot [color=red, only marks, mark=o, mark options={solid, red}, mark size=4pt]
		table[row sep=crcr]{%
			0.38036099394506	-0.0323020684174378\\
			0.0323020684174378	0.38036099394506\\
			-0.38036099394506	0.0323020684174378\\
			-0.0323020684174378	-0.38036099394506\\
		};
	\end{axis}
\end{tikzpicture}%
		\end{subfigure}
	\end{minipage}
	\caption{One visual example of the clustering process in \eqref{equ_simplified}, where DQPSK is considered. The upper left figure shows the mean values $\{e^{-jS_l}\hat{r}_{i,m'}\}$, and the lower left figure shows the mean values $\{e^{-jS_q}\hat{r}_{i,m'+1}\}$. In the right figure, the green points represent the overall mean values $\{\hat{\nu}_{i,m'+1,l,q}\}$ in \eqref{equ_exactmultiplication} and the red points represent the approximated mean values  $\{e^{j(l-1)\frac{2\pi}{V}}\hat{\nu}_{i,m'+1}\}$ in \eqref{equ_simplified}}
	\label{fig_clustering}
\end{figure}

Assume $|\hat{r}_{i,m'}|>|\hat{r}_{i,m'+1}|$. Then $\frac{1}{V}\mathcal{CN}(\nu_{i};e^{-jS_l}\hat{r}_{i,m'},v^r_{i,m'})$ ($S_l\in\{S_1,\cdots,S_V\}$) is multiplied with $\sum_{q=1}^{V}\frac{1}{V}\mathcal{CN}(\nu_{i};e^{-jS_q}\hat{r}_{i,m'+1},v^r_{i,m'+1})$ to obtain one group of Gaussian functions $\sum_{q=1}^{V}\xi_{l,q}\mathcal{CN}(\nu_{i};\hat{\nu}_{i,m'+1,l,q},v^{\nu}_{i,m'+1})$. Approximate $\sum_{q=1}^{V}\xi_{l,q}\mathcal{CN}(\nu_{i};\hat{\nu}_{i,m'+1,l,q},v^{\nu}_{i,m'+1})$ as $\xi_{l}\mathcal{CN}(\nu_{i};\hat{\nu}_{i,m'+1,l},v^{\nu}_{i,m'+1,l})$, where the approximated mean $\hat{\nu}_{i,m'+1,l}$, variance $v^{\nu}_{i,m'+1,l}$ and weight $\xi_{l}$ are respectively calculated as
\begin{subequations}
	\begin{align}
		\hat{\nu}_{i,m'+1,l}&=\int_{-\infty}^{\infty}\nu_{i}\sum_{q=1}^{V}\xi_{l,q}\mathcal{CN}(\nu_{i};\hat{\nu}_{i,m'+1,l,q},v^{\nu}_{i,m'+1})d\nu_{i}
		=\sum_{q=1}^{V}\xi_{l,q}\hat{\nu}_{i,m'+1,l,q}\\
		v^{\nu}_{i,m'+1,l}&=\int_{-\infty}^{\infty}\left(\nu_{i}-\hat{\nu}_{i,m'+1,l}\right)^2\sum_{q=1}^{V}\xi_{l,q}\mathcal{CN}(\nu_{i};\hat{\nu}_{i,m'+1,l,q},v^{\nu}_{i,m'+1})d\nu_{i}
		\\&\nonumber=v^{\nu}_{i,m'+1}+\sum_{q=1}^{V}\xi_{l,q}\left(\hat{\nu}_{i,m'+1,l,q}-\hat{\nu}_{i,m'+1,l}\right)^2\\
		\xi_{l}&=\sum_{q=1}^{V}\xi_{l,q}.
	\end{align}
\end{subequations}
The approximated Gaussian function for the $l$-th group is $\xi_{l}\mathcal{CN}(\nu_{i};\hat{\nu}_{i,m'+1,l},v^{\nu}_{i,m'+1,l})$, and the approximated Gaussian functions for the rest groups can be directly obtained, where the mean values are calculated by adding additional phase shifts $q\frac{2\pi}{V}$ ($q\in\{1,\cdots,V-1\}$) on $\hat{\nu}_{i,m'+1,l}$ (i.e., $e^{jq\frac{2\pi}{V}}\hat{\nu}_{i,m'+1,l}$), and the variances are the same with $v^{\nu}_{i,m'+1,l}$. Dropping the common group index $l$ of $\hat{\nu}_{i,m'+1,l}$ and $v^{\nu}_{i,m'+1,l}$, \eqref{equ_exactmultiplication} is simplified as
\begin{equation}\label{equ_simplified}
	\begin{aligned}
		&\left(\sum_{l=1}^{V}\frac{1}{V}\mathcal{CN}(\nu_{i};e^{-jS_l}\hat{r}_{i,m'},v^{r}_{i,m'})\right)\left(\sum_{q=1}^{V}\frac{1}{V}\mathcal{CN}(\nu_{i};e^{-jS_q}\hat{r}_{i,m'+1},v^{r}_{i,m'+1})\right)\\\approx&
		\sum_{l=1}^{V}\xi_{l}\mathcal{CN}(\nu_{i};e^{j(l-1)\frac{2\pi}{V}}\hat{\nu}_{i,m'+1},v^{\nu}_{i,m'+1}).
	\end{aligned}
\end{equation} 
One visual example of the clustering process is shown in Fig. \ref{fig_clustering}.
When \eqref{equ_simplified} is further multiplied with the $(m'+2)$-th Gaussian mixture, \eqref{equ_exactmultiplication}-\eqref{equ_simplified} is repeated until all the $M-1$ Gaussian mixtures are multiplied together. 
Then \eqref{equ_mvp} is simplified as  
\begin{equation}\label{equ_mvpz}
	\begin{aligned}
		\mathcal{M}_{\nu_i\rightarrow p(\zeta_{i,m}|\nu_i,\chi_{i,m})}(\nu_i)&\approx\left((1-\rho)\delta(\nu_i)+\rho\mathcal{CN}(\nu_i;\hat{\beta},v^{\beta})\right)\sum_{l=1}^{V}\xi_{l}\mathcal{CN}(\nu_{i};e^{j(l-1)\frac{2\pi}{V}}\hat{\nu}_{i,M-1},v^{\nu}_{i,M-1})\\&=\left(1-\pi_{\nu_i\rightarrow p(\zeta_{i,m}|\nu_i,\chi_{i,m})}\right)\delta(\nu_i)+\pi_{\nu_i\rightarrow p(\zeta_{i,m}|\nu_i,\chi_{i,m})}\sum_{l=1}^{V}\xi'_l\mathcal{CN}(\nu_{i};\hat{\nu}_{i,l},v^{\nu}_{i}),
	\end{aligned}
\end{equation}
where $\hat{\nu}_{i,M-1}$ and $v^{\nu}_{i,M-1}$ are the approximated mean and variance of the products of $M-1$ Gaussian mixtures and 
\begin{subequations}
	\begin{align}
		v^{\nu}_{i} &= \left(\frac{1}{v^{\nu}_{i,M-1}}+\frac{1}{v^{\beta}}\right)^{-1}\\
		\hat{\nu}_{i,l} &= v^{\nu}_{i} \left(\frac{e^{j(l-1)\frac{2\pi}{V}}\hat{\nu}_{i,M-1}}{v^{\nu}_{i,M-1}}+\frac{\hat{\beta}}{v^{\beta}}\right)\\
		\xi'_l &= \frac{\xi_l\mathcal{CN}(0;\hat{\beta}-e^{j(l-1)\frac{2\pi}{V}}\hat{\nu}_{i,M-1},v^{\beta}+v^{\nu}_{i,M-1})}{\sum_{q=1}^{V}\xi_{q}\mathcal{CN}(0;\hat{\beta}-e^{j(q-1)\frac{2\pi}{V}}\hat{\nu}_{i,M-1},v^{\beta}+v^{\nu}_{i,M-1})}\\
		\pi_{\nu_i\rightarrow p(\zeta_{i,m}|\nu_i,\chi_{i,m})}&=\left(1+\frac{(1-\rho)\sum_{q=1}^{V}\mathcal{CN}(0;e^{j(q-1)\frac{2\pi}{V}}\hat{\nu}_{i,M-1},v^{\nu}_{i,M-1})}{\rho\sum_{q=1}^{V}\xi_{q}\mathcal{CN}(0;\hat{\beta}-e^{j(q-1)\frac{2\pi}{V}}\hat{\nu}_{i,M-1},v^{\beta}+v^{\nu}_{i,M-1})}\right)^{-1}.
	\end{align}
\end{subequations}

With the prior probability \eqref{equ_chi} and the Bernoulli Gaussian mixture \eqref{equ_mvpz}, the message from factor node $p(\zeta_{i,m}|\nu_i,\chi_{i,m})$ to variable node $\zeta_{i,m}$ is calculated as 
\begin{equation}\label{equ_message}
	\begin{aligned}
		\mathcal{M}_{p(\zeta_{i,m}|\nu_i,\chi_{i,m})\rightarrow \zeta_{i,m}}(\zeta_{i,m})&=\int_{\nu_i}\sum_{\chi_{i,m}}p(\zeta_{i,m}|\nu_i,\chi_{i,m})\mathcal{M}_{\nu_i\rightarrow p(\zeta_{i,m}|\nu_i,\chi_{i,m})}(\nu_i)\mathcal{M}_{\chi_{i,m}\rightarrow p(\zeta_{i,m}|\nu_i,\chi_{i,m})}(\chi_{i,m})
		\\&=\int_{\nu_i}\sum_{\chi_{i,m}}\delta(\zeta_{i,m}-\nu_i\chi_{i,m})\mathcal{M}_{\nu_i\rightarrow p(\zeta_{i,m}|\nu_i,\chi_{i,m})}(\nu_i)\sum_{q=1}^{V}\frac{1}{V}\delta(\chi_{i,m}-e^{jS_q})
		\\&=\left(1-\pi_{\nu_i\rightarrow p(\zeta_{i,m}|\nu_i,\chi_{i,m})}\right)\delta(\zeta_{i,m})+\pi_{\nu_i\rightarrow p(\zeta_{i,m}|\nu_i,\chi_{i,m})}\sum_{l=1}^{V}\sum_{q=1}^{V}\frac{\xi'_l}{V}\mathcal{CN}(\zeta_{i,m};e^{jS_q}\hat{\nu}_{i,l},v^{\nu}_{i})
		\\&\overset{(a)}{\approx}\left(1-\pi_{\nu_i\rightarrow p(\zeta_{i,m}|\nu_i,\chi_{i,m})}\right)\delta(\zeta_{i,m})+\pi_{\nu_i\rightarrow p(\zeta_{i,m}|\nu_i,\chi_{i,m})}\sum_{l=1}^{V}\xi'_l\mathcal{CN}(\zeta_{i,m};\hat{\nu'}_{i,l},v^{\nu}_{i}),
	\end{aligned}
\end{equation}
where $(a)$ follows the proposed progressive approximation detailed in \eqref{equ_exactmultiplication}-\eqref{equ_simplified}, and $\hat{\nu'}_{i,l}$ is the approximated mean value for the $l$-th cluster.
\subsubsection{Message Update at $\zeta_{i,m}$}
By combining message $\mathcal{M}_{\zeta_{i,m}\rightarrow p(\zeta_{i,m}|\nu_i,\chi_{i,m})}(\zeta_{i,m})\sim \mathcal{CN}(\zeta_{i,m};\hat{r}_{i,m},v^r_{i,m})$ and \eqref{equ_message},
the belief of $\zeta_{i,m}$ is expressed as
\begin{equation}\label{equ_postzeta}
	\begin{aligned}
		\mathcal{M}(\zeta_{i,m})&=\mathcal{M}_{\zeta_{i,m}\rightarrow p(\zeta_{i,m}|\nu_i,\chi_{i,m})}(\zeta_{i,m})\mathcal{M}_{p(\zeta_{i,m}|\nu_i,\chi_{i,m})\rightarrow \zeta_{i,m}}(\zeta_{i,m})
		\\&=\mathcal{CN}(\zeta_{i,m};\hat{r}_{i,m},v^{r}_{i,m})\left(\left(1-\pi_{\nu_i\rightarrow p(\zeta_{i,m}|\nu_i,\chi_{i,m})}\right)\delta(\zeta_{i,m})+\pi_{\nu_i\rightarrow p(\zeta_{i,m}|\nu_i,\chi_{i,m})}\sum_{l=1}^{V}\xi'_l\mathcal{CN}(\zeta_{i,m};\hat{\nu'}_{i,l},v^{\nu}_{i})\right)
		\\&=(1-\pi_{\zeta_{i,m}})\delta(\zeta_{i,m})+\pi_{\zeta_{i,m}}\sum_{l=1}^{V}\xi_{\zeta_{i,m},l}\mathcal{CN}(\zeta_{i,m};\hat{\zeta}_{i,m,l},v^{\zeta}_{i,m}),
	\end{aligned}
\end{equation}
where the variance $v^{\zeta}_{i,m}$, the mean $\hat{\zeta}_{i,m,l}$, the weight $\xi_{\zeta_{i,m},l}$ and the probability $\pi_{\zeta_{i,m}}$ are respectively given by
\begin{subequations}
	\begin{align}
		v^{\zeta}_{i,m} &= \left(\frac{1}{v^{r}_{i,m}}+\frac{1}{v^{\nu}_{i}}\right)^{-1}\\
		\hat{\zeta}_{i,m,l} &= v^{\zeta}_{i,m}\left(\frac{\hat{r}_{i,m}}{v^{r}_{i,m}}+\frac{\hat{\nu'}_{i,l}}{v^{\nu}_{i}}\right)\\
		\xi_{\zeta_{i,m},l} &= \frac{\xi'_l\mathcal{CN}(0;\hat{r}_{i,m}-\hat{\nu'}_{i,l},v^{r}_{i,m}+v^{\nu}_{i})}{\sum_{q=1}^{V}\xi'_{q}\mathcal{CN}(0;\hat{r}_{i,m}-\hat{\nu'}_{i,q},v^{r}_{i,m}+v^{\nu}_{i})}\\
		\pi_{\zeta_{i,m}}&=\left(1+\frac{(1-\pi_{\nu_i\rightarrow p(\zeta_{i,m}|\nu_i,\chi_{i,m})})\mathcal{CN}(0;\hat{r}_{i,m},v^{r}_{i,m})}{\pi_{\nu_i\rightarrow p(\zeta_{i,m}|\nu_i,\chi_{i,m})}\sum_{l=1}^{V}\xi'_l\mathcal{CN}(0;\hat{r}_{i,m}-\hat{\nu'}_{i,l},v^{r}_{i,m}+v^{\nu}_{i})}\right)^{-1}.\label{equ_pizeta}
	\end{align}
\end{subequations}
With \eqref{equ_postzeta}, the mean and variance of $\zeta_{i,m}$ are respectively calculated as 
\begin{subequations}
	\begin{align}\label{equ_meanzetafinal}
		\hat{\zeta}_{i,m}&=\int_{-\infty}^{\infty}\zeta_{i,m}\pi_{\zeta_{i,m}}\sum_{l=1}^{V}\xi_{\zeta_{i,m},l}\mathcal{CN}(\zeta_{i,m};\hat{\zeta}_{i,m,l},v^{\zeta}_{i,m})d\zeta_{i,m}=\pi_{\zeta_{i,m}}\sum_{l=1}^{V}\xi_{\zeta_{i,m},l}\hat{\zeta}_{i,m,l}\\\label{equ_variancezetafinal}v^{\zeta}_{i,m}&=\int_{-\infty}^{\infty}(\zeta_{i,m}-\hat{\zeta}_{i,m})^2\left((1-\pi_{\zeta_{i,m}})\delta(\zeta_{i,m})+\pi_{\zeta_{i,m}}\sum_{l=1}^{V}\xi_{\zeta_{i,m},l}\mathcal{CN}(\zeta_{i,m};\hat{\zeta}_{i,m,l},v^{\zeta}_{i,m})\right)d\zeta_{i,m}
		\\&\nonumber=(1-\pi_{\zeta_{i,m}})\hat{\zeta}_{i,m}^2+\pi_{\zeta_{i,m}}\sum_{l=1}^{V}\xi_{\zeta_{i,m},l}\left(v^{\zeta}_{i,m}+(\hat{\zeta}_{i,m,l}-\hat{\zeta}_{i,m})^2\right).
	\end{align}
\end{subequations}

The overall algorithm is summarized in Algorithm \ref{alg_markov}, where steps 3-10 are based on the GAMP described in \cite{6556987}. Steps 3-4 are the output linear step to obtain the quantities $\{\hat{p}_{j,m}\}$ and $\{v^p_{j,m}\}$. Using these quantities, steps 5-6 compute the marginal posterior means $\{\hat{q}_{j,m}\}$ and variances $\{v^q_{j,m}\}$. Steps 7-8 then use these posterior moments to compute the scaled residuals $\{\hat{a}_{j,m}\}$ and the inverse residual variances $\{v^a_{j,m}\}$. Steps 9-10 correspond to the input linear step to compute $\{v^r_{i,m}\}$ and $\{\hat{r}_{i,m}\}$, where $\hat{r}_{i,m}$ can be interpreted as an observation of $\zeta_{i,m}$ under an additive white Gaussian noise (AWGN) channel with zero mean and a variance of $v^r_{i,m}$.
	
\begin{algorithm}[!htbp] 
	\caption{The Message-Passing Algorithm} 
	\label{alg_markov} 
	\begin{algorithmic}[1] 
			\Require 
			$\mathbf{Y}$, $\mathbf{Z}_1$, $\cdots$, $\mathbf{Z}_M$, prior distributions $p(\bm{\chi})$
			\Ensure 
			$\hat{\bm{\zeta}}$, $\mathbf{V}^{\bm{\zeta}}$
			\State Initialization: $\forall i,j,m:$ $\hat{\zeta}_{i,m}(1)$, $v^\zeta_{i,m}(1)$, $\hat{a}_{j,m}(0)=0$;
			\For{$l=1,\cdots$,$L_{\max}$}
			\State $\forall j,m: 
			{v}^p_{j,m}(l)=\sum_i|Z_{j,i,m}|^2v^\zeta_{i,m}(l)$
			\State $\forall j,m: \hat{p}_{j,m}(l)=\sum_iZ_{j,i,m}\hat{\zeta}_{i,m}(l)-{v}^p_{j,m}(l)\hat{a}_{j,m}(l-1)$
			\State $\forall j,m:  \hat{q}_{j,m}(l)=E\{\hat{q}_{j,m}|\hat{y}_{j,m},{v}^y_{j,m},\hat{p}_{j,m}(l),{v}^p_{j,m}(l)\}$
			\State $\forall j,m:  v^q_{j,m}(l)=Var\{\hat{q}_{j,m}|\hat{y}_{j,m},{v}^y_{j,m},\hat{p}_{j,m}(l),{v}^p_{j,m}(l)\}$
			\State $\forall j,m:\hat{a}_{j,m}(l) = (\hat{q}_{j,m}(l)-\hat{p}_{j,m}(l))/{v}^p_{j,m}(l)$
			\State $\forall j,m:v^a_{j,m}(l)= (1-v^q_{j,m}(l)/{v}^p_{j,m}(l))/{v}^p_{j,m}(l)$
			\State $\forall i,m:v^r_{i,m}(l)= (\sum_j|Z_{j,i,m}|^2v^a_{j,m}(l))^{-1}$
			\State $\forall i,m:\hat{r}_{i,m}(l) = \hat{\zeta}_{i,m}(l)+v^r_{i,m}(l)\sum_jZ_{j,i,m}\hat{a}_{j,m}(l)$
			\State $\forall i,m:$ Update $\mathcal{M}_{p(\zeta_{i,m}|\nu_i,\chi_{i,m})\rightarrow\nu_i}(\nu_i)$ by \eqref{equ_up}
			\State $\forall i,m:$ Update $\mathcal{M}_{\nu_i\rightarrow p(\zeta_{i,m}|\nu_i,\chi_{i,m})}(\nu_i)$ by \eqref{equ_mvpz}
			\State $\forall i,m:$ Update $\mathcal{M}_{p(\zeta_{i,m}|\nu_i,\chi_{i,m})\rightarrow \zeta_{i,m}}(\zeta_{i,m})$ by \eqref{equ_message}
			\State $\forall i,m:$ Update $\mathcal{M}(\zeta_{i,m})$ by \eqref{equ_postzeta}, and obtain $\hat{\zeta}_{i,m}(l+1)$ by \eqref{equ_meanzetafinal} and $v^\zeta_{i,m}(l+1)$ by \eqref{equ_variancezetafinal}
			\State \textbf{if} a stopping criterion is met, \textbf{stop}
			\EndFor
	\end{algorithmic}
\end{algorithm}

\subsection{Complexity Analysis}
The computational complexity of Algorithm \ref{alg_markov} mainly comes from step 11 due to the multiple multiplications of Gaussian mixtures. In specific,
there are $2V^{M-1}$ ($M\geq3$) multiplications of two complex Gaussian functions while calculating \eqref{equ_mvp}. However, with our proposed progressive approximation method, the number is reduced to $V+(M-1)V^2$, which largely reduces the computational cost. The multiplication between two complex Gaussian functions requires $\mathcal{O}(1)$ computational cost. Moreover, steps 3-10 require $\mathcal{O}(N_\mathrm{r}NUQM)$. Then the overall computational cost in Algorithm \ref{alg_markov} is $\mathcal{O}\left((N_\mathrm{r}N+V+(M-1)V^2)UQM\right)$ per iteration.

For comparisons, the complexity of three state-of-the-art algorithms, namely, Bernoulli-Gaussian generalized approximate message passing (BG-GAMP) \cite{6556987}, orthogonal matching pursuit (OMP) \cite{4385788}, turbo compressive sensing (Turbo-CS) \cite{6883198} and sparse Bayesian learning (SBL) \cite{tipping2001sparse}, are provided in Table \ref{tab_complexity}. In BG-GAMP, each element of $\bm{\zeta}$ has its own $\nu$ instead of a shared $\nu$ for each row in Algorithm \ref{alg_markov}, then steps 11-12 are ignored. In OMP, for obtaining the possible non-zero indices the main complexity comes form calculating the product between each column of $\mathbf{Z}$ and each column of $\mathbf{Y}$ resulting in $\mathcal{O}((N_\mathrm{r}N)^2M)$ per iteration. The main computational costs in both Turbo-CS and SBL lie in the matrix inversions and matrix multiplications, which are $\mathcal{O}((N_rN)^3)$ and $\mathcal{O}((N_\mathrm{r}N)^2UQM)$ per iteration, respectively.

\begin{table}[!htbp]
	\centering
	\caption{Complexity analysis}
	\begin{tabular}{ccc}
		\hline
		Algorithms           & Computational costs                                  \\ \hline
		BG-GAMP \cite{6556987}&  $\mathcal{O}(N_\mathrm{r}NUQM)$                       \\
		OMP \cite{4385788}&  $\mathcal{O}((N_\mathrm{r}N)^2M)$                       \\
		Turbo-CS \cite{6883198} & $\mathcal{O}(((N_\mathrm{r}N)^3+(N_\mathrm{r}N)^2UQ)M)$\\
		SBL \cite{tipping2001sparse} &  $\mathcal{O}(((N_\mathrm{r}N)^3+(N_\mathrm{r}N)^2UQ)M)$                       \\
		Algorithm \ref{alg_markov} &  $\mathcal{O}((N_\mathrm{r}N+V+(M-1)V^2)UQM)$                       \\
		\hline
	\end{tabular}\label{tab_complexity}
\end{table}

\section{Parameters Learning}
In practice, the prior parameters $\sigma^2$ and $\rho$ are usually unknown and required to be estimated. Moreover, the true localization parameters $\bm{\vartheta}$ and $\bm{\tau}$ may not be the grid values illustrated in \eqref{equ_grids}, and usually the values between adjacent grids resulting in the model mismatch problem. To address these issues, we treat $\omega=\{\sigma^2,\rho,\bm{\vartheta},\bm{\tau}\}$ as unknown parameters and utilize the EM approach \cite{543975} to learn these parameters for improving the proposed algorithm. The EM process is described as
\begin{equation}\label{equ_emall}
	\omega(l+1)=\arg\max_{\omega}E_{p(\bm{\zeta}|\mathbf{Y};\omega(l))}\{\ln p(\mathbf{Y},\bm{\zeta};\omega)\},
\end{equation}
where $\omega(l)=\{\sigma^2(l),\rho(l),\bm{\vartheta}(l),\bm{\tau}(l)\}$ is the estimate of $\omega$ in the $l$-th EM-learning iteration. $E_{p(\bm{\zeta}|\mathbf{Y};\omega(l))}\{\cdot\}$ represents the expectation over the posterior distribution $p(\bm{\zeta}|\mathbf{Y};\omega(l))$. Since it is impractical to update all parameters in $\omega$ at once, we update each parameter at a time while fixing the others.

\subsection{Learning $\sigma^2$ and $\rho$} \label{secparaleaning}
We first derive the EM update for the noise variance $\sigma^2$ given previous parameters $\omega(l)$. The joint probability density function (pdf) $p(\mathbf{Y},\bm{\zeta};\sigma^2)$ in \eqref{equ_emall} is decoupled as
\begin{equation}\label{equ_noise}
	{\sigma^2}(l+1)
	=\arg\max_{\sigma^2}\int_{\bm{\zeta}} p(\bm{\zeta}|\mathbf{Y};\omega(l))\ln p(\mathbf{Y}|\bm{\zeta};\sigma^2)+\int_{\bm{\zeta}} p(\bm{\zeta}|\mathbf{Y};\omega(l))\ln p(\bm{\zeta};\sigma^2).
\end{equation}
Setting the derivative of \eqref{equ_noise} (with respect to $\sigma^2$) to zero, yielding
\begin{equation}
	\int_{\bm{\zeta}} p(\bm{\zeta}|\mathbf{Y};\omega(l))\frac{\ln\partial p(\mathbf{Y}|\bm{\zeta};\sigma^2)}{\partial \sigma^2}=0,
\end{equation}
where the last term of \eqref{equ_noise} is dropped since it is irrelevant to $\sigma^2$ by noting $p(\bm{\zeta};\sigma^2)=p(\bm{\zeta})$. With \eqref{equ_yzeta}, the update of $\sigma^2$ is
\begin{equation}\label{equ_sigmalearning}
	{\sigma^2}(l+1)=\frac{1}{N_\mathrm{r}NM}\sum_{m=1}^{M}Tr\{(\mathbf{y}_m-\mathbf{Z}_m\hat{\bm{\zeta}}_{m}(l))(\mathbf{y}_m-\mathbf{Z}_m\hat{\bm{\zeta}}_{m}(l))^H+\mathbf{Z}_m\mathbf{V}^{\bm{\zeta}}_{m}(l)\mathbf{Z}_m^H\},
\end{equation} 
where $\hat{\bm{\zeta}}_{m}(l)$ and $\mathbf{V}^{\bm{\zeta}}_{m}(l)$ are the posterior mean and variance of $\bm{\zeta}$ at the $m$-th column in the $l$-th EM-learning iteration, and calculated by \eqref{equ_meanzetafinal} and \eqref{equ_variancezetafinal}, respectively.

Similar to \eqref{equ_noise}, the EM update for the sparsity $\rho$ can be written as
\begin{equation}\label{equ_rho}
	{\rho}(l+1)=\arg\max_{\rho}\int_{\bm{\zeta}} p(\bm{\zeta}|\mathbf{Y};\omega(l))\ln p(\mathbf{Y}|\bm{\zeta};\rho)+\int_{\bm{\zeta}} p(\bm{\zeta}|\mathbf{Y};\omega(l))\ln p(\bm{\zeta};\rho).
\end{equation}
We set the derivatives of \eqref{equ_rho} (with respect to $\rho$) to zero, which is
\begin{equation}
	\int_{\bm{\zeta}} p(\bm{\zeta}|\mathbf{Y};\omega(l))\frac{\ln\partial p(\bm{\zeta};\rho)}{\partial \rho}=0,
\end{equation}
where the first term is dropped since it is irrelevant to $\rho$.
With \eqref{equ_nu}, the update of $\rho$ is \cite{6556987}
\begin{equation}\label{equ_rhoup}
	\rho(l+1) = \frac{1}{UQM}\sum_{i=1}^{UQ}\sum_{m=1}^{M}\pi_{\zeta_{i,m}},
\end{equation}
where $\pi_{\zeta_{i,m}}$ is defined in \eqref{equ_pizeta}. 

\subsection{Learning $\bm{\vartheta}$ and $\bm{\tau}$}
The objective function in \eqref{equ_emall} is reduced to
\begin{equation}
	\omega(l+1) = \arg\max_{\omega} \int_{\bm{\zeta}}p(\bm{\zeta}|\mathbf{Y};\omega(l))\ln p(\mathbf{Y}|\bm{\zeta};\omega),
	\label{equ_max}
\end{equation}
where the term $\int_{\bm{\zeta}} p(\bm{\zeta}|\mathbf{Y};\omega(l))\ln p(\bm{\zeta};\omega)$ is dropped since it is irrelevant to $\omega$ by noting $p(\bm{\zeta};\omega)=p(\bm{\zeta})$. Inserting \eqref{equ_yzeta} into \eqref{equ_max}, we obtain
\begin{equation}
	\begin{aligned}
		\int_{\bm{\zeta}}p(\bm{\zeta}|\mathbf{Y};\omega(l))\ln p(\mathbf{Y}|\bm{\zeta};\omega) &= \sum_{m=1}^{M}
		\int_{\bm{\zeta}_m}p(\bm{\zeta}_m|\mathbf{y}_m;\omega(l))\ln\mathcal{CN}(\mathbf{y}_m;\mathbf{Z}_m\bm{\zeta}_m,\sigma^2\mathbf{I}_{N_\mathrm{r}N})\\&
		=-M\ln(\pi\sigma^2)-\frac{1}{\sigma^2}\sum_{m=1}^{M}\int_{\bm{\zeta}_m}p(\bm{\zeta}_m|\mathbf{y}_m;\omega(l))\\&
		\qquad\times(\mathbf{y}_m^H\mathbf{y}_m-\mathbf{y}_m^H\mathbf{Z}_m\bm{\zeta}_m-(\mathbf{Z}_m\bm{\zeta}_m)^H\mathbf{y}_m+(\mathbf{Z}_m\bm{\zeta}_m)^H(\mathbf{Z}_m\bm{\zeta}_m)).
	\end{aligned}
\end{equation}
By noting that $\mathbf{Z}_m$ is a function of $\omega$ and ignoring the terms independent of $\omega$, \eqref{equ_max} is reduced to
\begin{equation}\label{equ_optmax}
	\begin{aligned}
		&\arg\max_{\omega}\int_{\bm{\zeta}}p(\bm{\zeta}|\mathbf{Y};\omega(l))\ln p(\mathbf{Y}|\bm{\zeta};\omega) \\ =&\arg\max_{\omega}\sum_{m=1}^{M}\int_{\bm{\zeta}_m}\mathcal{CN}(\bm{\zeta}_m;\bm{\hat{\zeta}}_{m}(l),\mathbf{V}^{\bm{\zeta}}_{m}(l))(\mathbf{y}_m^H\mathbf{Z}_m\bm{\zeta}_m+(\mathbf{Z}_m\bm{\zeta}_m)^H\mathbf{y}_m-(\mathbf{Z}_m\bm{\zeta}_m)^H(\mathbf{Z}_m\bm{\zeta}_m))\\
		=&\arg\max_{\omega}\sum_{m=1}^{M}2Re\{\mathbf{y}_m^H\mathbf{Z}_m\hat{\bm{\zeta}}_{m}(l)\}-Tr\{\mathbf{Z}_m^H\mathbf{Z}_m\int_{\bm{\zeta}_m}\bm{\zeta}_m\bm{\zeta}_m^H\mathcal{CN}(\bm{\zeta}_m;\bm{\hat{\zeta}}_{m}(l),\mathbf{V}^{\bm{\zeta}}_{m}(l))\}\\
		=&\arg\max_{\omega}\sum_{m=1}^{M}2Re\{\mathbf{y}_m^H\mathbf{Z}_m\hat{\bm{\zeta}}_{m}(l)\}-Tr\{\mathbf{Z}_m^H\mathbf{Z}_m\omega_{m}(l)\},	
	\end{aligned}
\end{equation}
where $\omega_{m}(l)=\bm{\hat{\zeta}}_{m}(l)\bm{\hat{\zeta}}_{m}(l)^H+\mathbf{V}^{\bm{\zeta}}_{m}(l)$. However, it is not easy to obtain a closed-form solution to \eqref{equ_optmax} since the problem is highly non-linear in $\bm{\vartheta}$ and $\bm{\tau}$. Therefore, we search for a new $\omega(l+1)$ via gradient descent as
\begin{equation}
	\omega(l+1) = \omega(l)+\epsilon\frac{\partial \mathcal{G}(\omega)}{\partial \omega},
	\label{equ_gradiant}
\end{equation}
where $\mathcal{G}(\omega)=\sum_{m=1}^M2Re\{\mathbf{y}_m^H\mathbf{Z}_m\hat{\bm{\zeta}}_{m}(l)\}-Tr\{\mathbf{Z}_m^H\mathbf{Z}_m\bm{\omega}_{m}(l)\}$; $\epsilon$ is an appropriate stepsize that is selected from the backtracking line search method, which constantly decreases the value of $\epsilon$ until $\omega(l+1)$ is found to make $\mathcal{G}(\omega(l+1))>\mathcal{G}(\omega(l))$. The gradient of $\mathcal{G}(\omega)$ with respect to $\vartheta_q$ and $\tau_u$ are respectively given by
\begin{subequations}\label{equ_deriva}
	\begin{equation}
		\begin{aligned}
			\frac{\partial \mathcal{G}(\omega)}{\partial \vartheta_q}
			=\sum_{m=1}^{M}2Re\left\{\mathbf{y}_m^H\frac{\partial \mathbf{Z}_m}{\partial \vartheta_q}\bm{\hat{\zeta}}_{m}(l)\right\}-Tr\left\{\left(\frac{\partial \mathbf{Z}_m^H}{\partial \vartheta_q}\mathbf{Z}_m+\mathbf{Z}_m^H\frac{\partial \mathbf{Z}_m}{\partial \vartheta_q}\right)\bm{\omega}_{m}(l)\right\}
		\end{aligned}
	\end{equation}
	\begin{equation}
		\begin{aligned}
			\frac{\partial \mathcal{G}(\omega)}{\partial \tau_u}
			=\sum_{m=1}^{M}2Re\left\{\mathbf{y}_m^H\frac{\partial \mathbf{Z}_m}{\partial \tau_u}\bm{\hat{\zeta}}_{m}(l)\right\}-Tr\left\{\left(\frac{\partial \mathbf{Z}_m^H}{\partial \tau_u}\mathbf{Z}_m+\mathbf{Z}_m^H\frac{\partial \mathbf{Z}_m}{\partial \tau_u}\right)\bm{\omega}_{m}(l)\right\}.
		\end{aligned}
	\end{equation}
\end{subequations}
From \eqref{equ_Z}, the derivatives of $\mathbf{Z}_m$ at the $n$-th subcarrier are respectively given by
\begin{subequations}\label{equ_deri}
	\begin{equation}
		\frac{\partial \mathbf{Z}_m[n]}{\partial \vartheta_q}= (e^{-j2\pi n\Delta f\bm{\tau}})^T \otimes \frac{\partial \mathbf{H}(\bm{\vartheta})}{\partial \vartheta_q}(\mathbf{I}_Q\otimes\mathbf{x}_m[n])
	\end{equation}
	\begin{equation}
		\frac{\partial \mathbf{Z}_m[n]}{\partial \tau_u}=\frac{\partial (e^{-j2\pi n\Delta f\bm{\tau}})^T}{\partial \tau_u} \otimes \mathbf{H}(\bm{\vartheta})(\mathbf{I}_Q\otimes\mathbf{x}_m[n]),
	\end{equation}
\end{subequations}
where
\begin{align}\nonumber
\frac{\partial \mathbf{H}(\bm{\vartheta})}{\partial \vartheta_q}&=\begin{bmatrix}
	\mathbf{0}_{N_\mathrm{r}\times N_\mathrm{t}}&\cdots&\mathbf{0}_{N_\mathrm{r}\times N_\mathrm{t}}&\frac{\partial \mathbf{H}(\vartheta_q)}{\partial \vartheta_q}&\mathbf{0}_{N_\mathrm{r}\times N_\mathrm{t}}&\cdots&\mathbf{0}_{N_\mathrm{r}\times N_\mathrm{t}}
\end{bmatrix}\\\nonumber
\frac{\partial (e^{-j2\pi n\Delta f\bm{\tau}})^T}{\partial \tau_u}&=\begin{bmatrix}
	0&\cdots&0&-j2\pi n\Delta fe^{-j2\pi n\Delta f\tau_u}&0&\cdots&0
\end{bmatrix}^T.
\end{align}

The overall EM-based parameters learning procedure is summarized in Algorithm \ref{alg1}. The computational complexity of Algorithm \ref{alg1} mainly lie in steps 8-9. Since we use the gradient descent method to iteratively maximize \eqref{equ_optmax}, the calculation of the gradient for each localization parameter in \eqref{equ_deriva} requires $\mathcal{O}(((UQ)^3+(UQ)^2N_\mathrm{r}N+UQN_\mathrm{r}N)M)$ per EM-learning iteration.
\begin{algorithm}[!htbp] 
	\caption{The EM-based Parameters Learning}
	\label{alg1}
	\begin{algorithmic}[1] 
		\Require $\mathbf{Y}$, prior distributions $p(\bm{\chi})$
		\Ensure 
		$\hat{\bm{\zeta}}$, $\mathbf{V}^{\bm{\zeta}}$, $\hat{\bm{\vartheta}}$, $\hat{\bm{\tau}}$, $\hat{\sigma^2}$ and $\hat{\rho}$
		\State Initialization: $\bm{\vartheta}(1)$, $\bm{\tau}(1)$, $\sigma^2(1)$, $\rho(1)$;
		\For{l=1,$\cdots$,$L_{\max}$}
		\For{$l'=1,\cdots$,$L'_{\max}$}
		\State Use the Algorithm \ref{alg_markov} to obtain $\bm{\hat{\zeta}}(l')$, $\mathbf{V}^{\bm{\zeta}}(l')$ and $\pi_{\zeta_{i,m}}(l')$.
		\State Update $\sigma^2(l'+1)$ by \eqref{equ_sigmalearning} and $\rho(l'+1)$ by \eqref{equ_rhoup}
		\State \textbf{if} a stopping criterion is met, \textbf{stop}
		\EndFor
		\State Fix $\bm{\tau}$, and maximize \eqref{equ_optmax} with respect to $\bm{\vartheta}$ through gradient descent \eqref{equ_gradiant} to obtain $\bm{\vartheta}(l+1)$.
		\State Fix $\bm{\vartheta}$, and maximize \eqref{equ_optmax} with respect to $\bm{\tau}$ through gradient descent \eqref{equ_gradiant} to obtain $\bm{\tau}(l+1)$.
		\State Generate new $\mathbf{Z}_m$ based on $\bm{\vartheta}(l+1)$ and $\bm{\tau}(l+1)$ by \eqref{equ_sysmode}.
		\State \textbf{if} a stopping criterion is met, \textbf{stop}
		\EndFor
	\end{algorithmic}
\end{algorithm}

\section{Bayesian Cram\'er-Rao Bound}\label{sec4}
We now develop a mean square error lower-bound of the considered estimation problem by assuming that the support of $\bm{\zeta}$ and the noise variance $\sigma^2$ are known. Then, the signal model in \eqref{equ_sysmode} reduces to 
\begin{equation}
	\mathbf{Y}=[\mathbf{Z}_{\mathrm{real},1}\bm{\zeta}_{\mathrm{real},1},\cdots,\mathbf{Z}_{\mathrm{real},M}\bm{\zeta}_{\mathrm{real},M}] + \mathbf{W},
	\label{equ_modsys2}
\end{equation}
where $\mathbf{Z}_{\mathrm{real},m}\in\mathbb{C}^{N_\mathrm{r}N\times K}$ is similar to $\mathbf{Z}_m$ but constructed by the angles $\bm{\vartheta}_{\mathrm{real}}=[\vartheta_1,\cdots,\vartheta_K]^T\in\mathbb{R}^{K\times 1}$ and delays $\bm{\tau}_{\mathrm{real}}=[\tau_1,\cdots,\tau_K]^T\in\mathbb{R}^{K\times 1}$; $\bm{\zeta}_{\mathrm{real},m}=[\alpha_{1,m},\cdots,\alpha_{K,m}]^T\in\mathbb{C}^{K\times 1}$ . By stacking the columns of \eqref{equ_modsys2} sequentially on top of one another, we have 
\begin{equation}
	\mathbf{y}=\mathbf{Z}_{\mathrm{real}}\bm{\zeta'}_{\mathrm{real}} + \mathbf{w},
\end{equation}
where
\begin{equation}\nonumber
	\mathbf{Z}_{\mathrm{real}}=\begin{bmatrix}\mathbf{Z}_{\mathrm{real},1}&\mathbf{0}&\cdots&\mathbf{0}\\ \mathbf{0}&\mathbf{Z}_{\mathrm{real},2}&\cdots&\mathbf{0}\\\vdots&\vdots&\cdots&\vdots\\\mathbf{0}&\mathbf{0}&\cdots&\mathbf{Z}_{\mathrm{real},M}\end{bmatrix},
\end{equation} $\bm{\zeta'}_{\mathrm{real}}=[\bm{\zeta}^T_{\mathrm{real},1},\bm{\zeta}^T_{\mathrm{real},2},\cdots,\bm{\zeta}^T_{\mathrm{real},M}]^T\in\mathbb{C}^{KM\times 1}$ and $\mathbf{w}\in\mathbb{C}^{N_\mathrm{r}NM\times 1}$. Let $\bm{\kappa}=[\bm{\vartheta}_{\mathrm{real}}^T, \bm{\tau}_{\mathrm{real}}^T, {\bm{\zeta'}}^T_{\mathrm{real}}]^T$ be the parameter vector. The Bayesian information matrix (BIM) \cite{Trees} is defined as
\begin{equation}
	\begin{aligned}
		\mathbf{J}_{\mathbf{y}}(\bm{\kappa})=\mathbf{J}^\mathrm{D}_{\mathbf{y}}(\bm{\kappa})+\mathbf{J}^\mathrm{P}_{\mathbf{y}}(\bm{\kappa}),
	\end{aligned}
	\label{equ_bayesian}
\end{equation}
where 
\begin{equation}\label{equ_dform}
	\mathbf{J}^\mathrm{D}_{\mathbf{y}}(\bm{\kappa}) = \begin{bmatrix}
		\mathbf{J}^\mathrm{D}_{\mathbf{y}}(\bm{\vartheta}_{\mathrm{real}}) & \mathbf{J}^\mathrm{D}_{\mathbf{y}}(\bm{\vartheta}_{\mathrm{real}},\bm{\tau}_{\mathrm{real}}) &
		\mathbf{J}^\mathrm{D}_{\mathbf{y}}(\bm{\vartheta}_{\mathrm{real}},\bm{\zeta'}_{\mathrm{real}}) \\
		\mathbf{J}^\mathrm{D}_{\mathbf{y}}(\bm{\tau}_{\mathrm{real}},\bm{\vartheta}_{\mathrm{real}}) & \mathbf{J}^\mathrm{D}_{\mathbf{y}}(\bm{\tau}_{\mathrm{real}}) &
		\mathbf{J}^\mathrm{D}_{\mathbf{y}}(\bm{\tau}_{\mathrm{real}},\bm{\zeta'}_{\mathrm{real}}) \\
		\mathbf{J}^\mathrm{D}_{\mathbf{y}}(\bm{\zeta'}_{\mathrm{real}},\bm{\vartheta}_{\mathrm{real}}) & \mathbf{J}^\mathrm{D}_{\mathbf{y}}(\bm{\zeta'}_{\mathrm{real}},\bm{\tau}_{\mathrm{real}}) &
		\mathbf{J}^\mathrm{D}_{\mathbf{y}}(\bm{\zeta'}_{\mathrm{real}})
	\end{bmatrix}
\end{equation}
and $\mathbf{J}^\mathrm{P}_{\mathbf{y}}(\bm{\kappa})$ has the similar form. Specifically, the $(i,l)$-th element of $\mathbf{J}^\mathrm{D}_{\mathbf{y}}(\bm{\kappa})$ and $\mathbf{J}^\mathrm{P}_{\mathbf{y}}(\bm{\kappa})$ are respectively calculated as
\begin{equation}
	[\mathbf{J}^\mathrm{D}_{\mathbf{y}}(\bm{\kappa})]_{i,l}=E_{\mathbf{y}|\bm{\kappa}}\left\{\frac{\partial \ln p(\mathbf{y}|\bm{\kappa})}{\partial \kappa_i}\frac{\partial \ln p(\mathbf{y}|\bm{\kappa})}{\partial \kappa_j}\right\}
	\label{equ_bayesianjd}
\end{equation}
\begin{equation}
	[\mathbf{J}^\mathrm{P}_{\mathbf{y}}(\bm{\kappa})]_{i,j} = E_{\bm{\kappa}}\left\{\frac{\partial \ln p(\bm{\kappa})}{\partial \kappa_i}\frac{\partial \ln p(\bm{\kappa})}{\partial \kappa_j}\right\}
	\label{equ_bayesianjp}
\end{equation}
with $\kappa_i$ being the $i$-th element of $\bm{\kappa}$. Since $\mathbf{y}\sim \mathcal{CN}(\mathbf{Z}_{\mathrm{real}}\bm{\zeta'}_{\mathrm{real}},\sigma^2\mathbf{I}_{N_\mathrm{r}NM})$, \eqref{equ_bayesianjd} is calculated as
\begin{equation}\label{equ_dkappa}
	[\mathbf{J}^\mathrm{D}_{\mathbf{y}}(\bm{\kappa})]_{i,j} = \frac{2}{\sigma^2}Re\left\{\left(\frac{\partial\mathbf{Z}_{\mathrm{real}}\bm{\zeta'}_{\mathrm{real}}}{\partial \kappa_i}\right)^H\frac{\partial\mathbf{Z}_{\mathrm{real}}\bm{\zeta'}_{\mathrm{real}}}{\partial \kappa_j}\right\}.
\end{equation} 
Based on \eqref{equ_dkappa}, the submatrices in \eqref{equ_dform} are 
\begin{subequations}\label{equ_dericrb}
	\begin{equation}\label{equ_de1}
		[\mathbf{J}^\mathrm{D}_{\mathbf{y}}(\bm{\vartheta}_{\mathrm{real}})]_{i,j} = \frac{2}{\sigma^2}Re\left\{{\bm{\zeta'}}^H_{\mathrm{real}}\frac{\partial\mathbf{Z}^H_{\mathrm{real}}}{\partial \vartheta_i}\frac{\partial\mathbf{Z}_{\mathrm{real}}}{\partial \vartheta_j}\bm{\zeta'}_{\mathrm{real}}\right\}
	\end{equation}
	\begin{equation}\label{equ_de2}
		[\mathbf{J}^\mathrm{D}_{\mathbf{y}}(\bm{\tau}_{\mathrm{real}})]_{i,j} = \frac{2}{\sigma^2}Re\left\{{\bm{\zeta'}}^H_{\mathrm{real}}\frac{\partial\mathbf{Z}^H_{\mathrm{real}}}{\partial \tau_i}\frac{\partial\mathbf{Z}_{\mathrm{real}}}{\partial \tau_j}\bm{\zeta'}_{\mathrm{real}}\right\}
	\end{equation}
\begin{equation}
	[\mathbf{J}^\mathrm{D}_{\mathbf{y}}(\bm{\zeta'}_{\mathrm{real}})]_{i,j} = \frac{2}{\sigma^2}Re\left\{\mathbf{e}^H_i\mathbf{Z}^H_{\mathrm{real}}\mathbf{Z}_{\mathrm{real}}\mathbf{e}_j\right\}
\end{equation}
	\begin{equation}\label{equ_deoff1}
		[\mathbf{J}^\mathrm{D}_{\mathbf{y}}(\bm{\vartheta}_{\mathrm{real}},\bm{\tau}_{\mathrm{real}})]_{i,j} = \frac{2}{\sigma^2}Re\left\{{\bm{\zeta'}}^H_{\mathrm{real}}\frac{\partial\mathbf{Z}^H_{\mathrm{real}}}{\partial \vartheta_i}\frac{\partial\mathbf{Z}_{\mathrm{real}}}{\partial \tau_j}\bm{\zeta'}_{\mathrm{real}}\right\}
	\end{equation}
	\begin{equation}\label{equ_deoff2}
		[\mathbf{J}^\mathrm{D}_{\mathbf{y}}(\bm{\vartheta}_{\mathrm{real}},\bm{\zeta'}_{\mathrm{real}})]_{i,j} = \frac{2}{\sigma^2}Re\left\{{\bm{\zeta'}}^H_{\mathrm{real}}\frac{\partial\mathbf{Z}^H_{\mathrm{real}}}{\partial \vartheta_i}\mathbf{Z}_{\mathrm{real}}\mathbf{e}_j\right\}
	\end{equation}
	\begin{equation}\label{equ_deoff3}
		[\mathbf{J}^\mathrm{D}_{\mathbf{y}}(\bm{\tau}_{\mathrm{real}},\bm{\zeta'}_{\mathrm{real}})]_{i,j} = \frac{2}{\sigma^2}Re\left\{{\bm{\zeta'}}^H_{\mathrm{real}}\frac{\partial\mathbf{Z}^H_{\mathrm{real}}}{\partial \tau_i}\mathbf{Z}_{\mathrm{real}}\mathbf{e}_j\right\},
	\end{equation}
\end{subequations}
where $\mathbf{e}_i$ and $\mathbf{e}_j$ are all-zero vectors except that the $i$-th and $j$-th element is one;  $\mathbf{J}^\mathrm{D}_{\mathbf{y}}(\bm{\tau}_{\mathrm{real}},\bm{\vartheta}_{\mathrm{real}})=\mathbf{J}^\mathrm{D}_{\mathbf{y}}(\bm{\vartheta}_{\mathrm{real}},\bm{\tau}_{\mathrm{real}})^H$, $\mathbf{J}^\mathrm{D}_{\mathbf{y}}(\bm{\zeta'}_{\mathrm{real}},\bm{\vartheta}_{\mathrm{real}})=\mathbf{J}^\mathrm{D}_{\mathbf{y}}(\bm{\vartheta}_{\mathrm{real}},\bm{\zeta'}_{\mathrm{real}})^H$, $\mathbf{J}^\mathrm{D}_{\mathbf{y}}(\bm{\zeta'}_{\mathrm{real}},\bm{\tau}_{\mathrm{real}})=\mathbf{J}^\mathrm{D}_{\mathbf{y}}(\bm{\tau}_{\mathrm{real}},\bm{\zeta'}_{\mathrm{real}})^H$. The involved derivatives can be found in \eqref{equ_deri}. Moreover, since $p(\bm{\zeta}_{\mathrm{real}})$ needs to be \emph{a priori} known in the Algorithm \ref{alg_markov}, we approximate $p(\zeta'_{\mathrm{real}})$ as a Gaussian mixture  $\sum_{l=1}^{V}\frac{1}{V}\mathcal{CN}(\zeta'_{\mathrm{real}};\hat{\beta}e^{jS_l},v^{\beta})$. The first derivative of the log-likelihood function $\ln p(\zeta'_{\mathrm{real}})$ is 
\begin{equation}
	\frac{\partial \ln p(\zeta'_{\mathrm{real}})}{\partial \zeta'_{\mathrm{real}}} = \frac{\sum_{l=1}^{V}e^{-\frac{(\zeta'_{\mathrm{real}}-\hat{\beta}e^{jS_l})^2}{v^{\beta}}}(-\frac{2(\zeta'_{\mathrm{real}}-\hat{\beta}e^{jS_l})}{v^{\beta}})}{\sum_{l=1}^{V}e^{-\frac{(\zeta'_{\mathrm{real}}-\hat{\beta}e^{jS_l})^2}{v^{\beta}}}},
\end{equation}
which can be inserted into \eqref{equ_bayesianjp} to obtain $[\mathbf{J}^\mathrm{P}_{\mathbf{y}}(\bm{\zeta'}_{\mathrm{real}})]_{i,j}$. Specifically, when $\hat{\beta}=0$, $[\mathbf{J}^\mathrm{P}_{\mathbf{y}}(\bm{\zeta'}_{\mathrm{real}})]_{i,j} =\frac{2}{v^{\beta}}$.
Since $\bm{\vartheta}_{\mathrm{real}}$ and $\bm{\tau}_{\mathrm{real}}$ are uniformly distributed, $[\mathbf{J}^\mathrm{P}_{\mathbf{y}}(\bm{\vartheta}_{\mathrm{real}})]_{i,j}$, $[\mathbf{J}^\mathrm{P}_{\mathbf{y}}(\bm{\tau}_{\mathrm{real}})]_{i,j}$, $[\mathbf{J}^\mathrm{P}_{\mathbf{y}}(\bm{\tau}_{\mathrm{real}},\bm{\vartheta}_{\mathrm{real}})]_{i,j}$, $[\mathbf{J}^\mathrm{P}_{\mathbf{y}}(\bm{\zeta'}_{\mathrm{real}},\bm{\vartheta}_{\mathrm{real}})]_{i,j}$ and $[\mathbf{J}^\mathrm{P}_{\mathbf{y}}(\bm{\zeta'}_{\mathrm{real}},\bm{\tau}_{\mathrm{real}})]_{i,j}$ are all zeros.
With \eqref{equ_bayesian}, the BCRB for the estimation of $\bm{\kappa}$ is given by
\begin{equation}
	Var\{\kappa_{i}\}=[\mathbf{J}^{-1}_\mathbf{y}(\bm{\kappa})]_{i,i}.
\end{equation}
Furthermore, we note that $\mathbf{d}_{\mathrm{real}} = \frac{c}{2}\bm{\tau}_{\mathrm{real}}$, 
where $\mathbf{d}_{\mathrm{real}}=[d_1,\cdots,d_K]^T\in\mathbb{R}^{K\times 1}$ are the distances between the BS and the devices with $c$ being the light speed. Thus, the BCRB for the estimation of $\mathbf{d}_{\mathrm{real}}$ is given by
\begin{equation}
	Var\{d_{{\mathrm{real}},i}\}=\frac{c^2}{4}Var\{\tau_{{\mathrm{real}},i}\}.
\end{equation}

\section{Numerical Results}\label{sec5}
The parameters used in the simulation are listed in Tab. \ref{tab_para}. The fading coefficient $\beta_k$ is given by \cite{9206044}
\begin{equation}
	\begin{aligned}
		\beta_k = \eta_k\sqrt{G_\mathrm{t}G_\mathrm{r}G\frac{L^2\frac{\lambda^4}{4}}{64\pi^3d_k^4}}
	\end{aligned}
	\label{equ_plk}
\end{equation}
with $\eta_k\sim\mathcal{CN}(0,1)$, $G_\mathrm{t}=100$, $G_\mathrm{r}=100$ and $G=1$ being the transmitter, receiver and RIS gains, respectively. Since the waveform propagates much faster than the devices' velocities, we make the approximation that the distances from the BS to the RIS and from the RIS to the BS are the same and denoted by $d_k$. 
Assuming that $d_k$ follows the uniform distribution within $[d_{\min},d_{\max}]$, the mean and variance for $d_k$ are calculated as $\frac{P_{\max}+P_{\min}}{2}$ and $\frac{P_{\min}^2+P_{\min} P_{\max}+P_{\max}^2}{3}$, respectively, where $P_{\max} = \sqrt{G_\mathrm{t}G_\mathrm{r}G\frac{L^2\frac{\lambda^4}{4}}{64\pi^3d_{\min}^4}}$ and $P_{\min} = \sqrt{G_\mathrm{t}G_\mathrm{r}G\frac{L^2\frac{\lambda^4}{4}}{64\pi^3d_{\max}^4}}$. Since $\eta_k$ and $d_k$ are two independent variables, the mean and variance of $\beta_k$ for \eqref{equ_nu} are calculated as
\begin{equation}
	\hat{\beta} = 0 \quad\text{and}\quad
	v^{\beta} = \frac{P_{\min}^2+P_{\min} P_{\max}+P_{\max}^2}{3}+\frac{(P_{\min}+P_{\max})^2}{4}.
\end{equation}
Moreover, the SNR is defined as $10\log{\frac{Tr\{(\mathbf{Z}_{\mathrm{real}}\bm{\zeta}_{\mathrm{real}})^H(\mathbf{Z}_{\mathrm{real}}\bm{\zeta}_{\mathrm{real}})\}}{N_\mathrm{r}NM\sigma^2}}$. 

\begin{table}[!htbp]
	\centering
	\caption{Global parameters}
	\begin{tabular}{ccc}
		\hline
		Parameters           & Symbols                & Values                    \\ \hline
		Subcarrier spacing&  $\Delta f$          & 15kHz                     \\
		CP duration&  $T_ \mathrm{cp}$                & 16.67$\mu$s \\
		Number of OFDM blocks&  $M$  & 20                    \\	
		Number of subcarriers &  $N$  & 30                     \\
		Number of transmit antennas&  $N_\mathrm{t}$ & 16                        \\
		Number of receive antennas&  $N_\mathrm{r}$  & 8                         \\
		Range of angle&  $\vartheta$       & [$-30^{\circ}$,$30^{\circ}$]           \\
		Number of reflecting elements on RIS&  $L$  & 30$\times$30          \\
		Number of devices&  $K$  & 5          \\\hline
	\end{tabular}\label{tab_para}
\end{table}
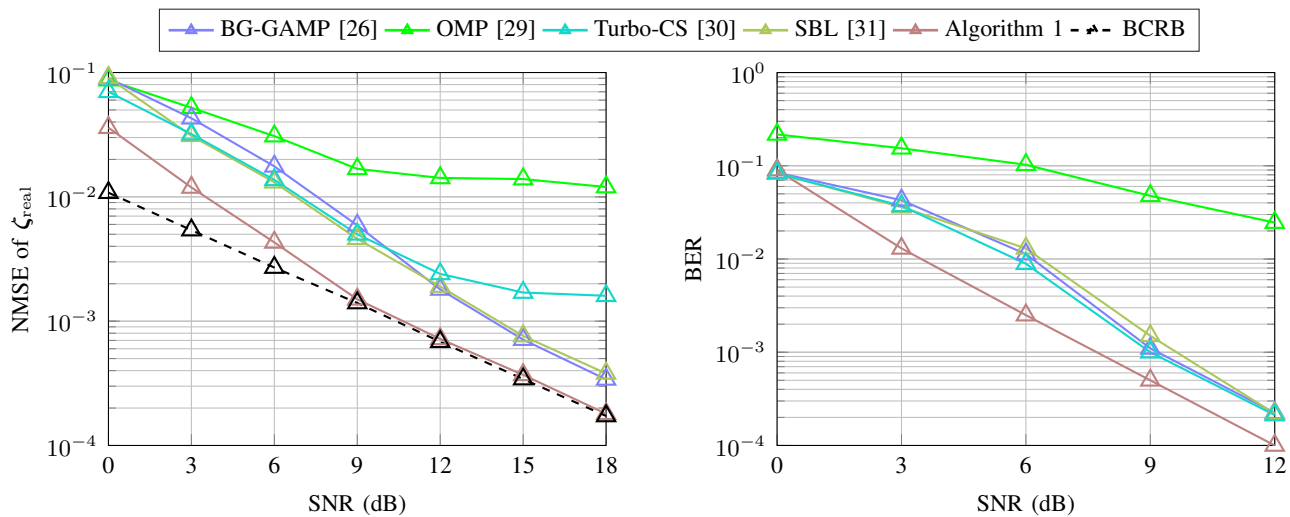
\begin{figure}[!htbp]
	\begin{minipage}{0.95\columnwidth}
		\hspace{2.1cm}\begin{subfigure}{0.95\columnwidth}
			\definecolor{clr1}{RGB}{128,0,0}
\definecolor{clr2}{RGB}{170,200,90}
\definecolor{clr5}{RGB}{17,215,205}

\begin{tikzpicture} 
	\begin{axis}[%
		hide axis,
		xmin=10,
		xmax=40,
		ymin=0,
		ymax=0.4,
		legend columns=6,
		legend style={draw=white!15!black,legend cell align=left,font=\footnotesize}
		]
		
		\addlegendimage{blue!50,line width=1.0pt,mark=triangle,mark size=2pt}
		\addlegendentry{BG-GAMP \cite{6556987}};
		
		\addlegendimage{green,line width=1.0pt,mark=triangle,mark size=2pt}
		\addlegendentry{OMP \cite{4385788}};
		
		\addlegendimage{clr5,line width=1.0pt,mark=triangle,mark size=2pt}
		\addlegendentry{Turbo-CS \cite{6883198}};
		
		\addlegendimage{clr2,line width=1.0pt,mark=triangle,mark size=2pt}
		\addlegendentry{SBL \cite{tipping2001sparse}};
		
		\addlegendimage{clr1!50,line width=1.0pt,mark=triangle,mark size=2pt}
		\addlegendentry{Algorithm \ref{alg_markov}};
		
		
		\addlegendimage{black,dashed,line width=1.0pt,mark=triangle,mark size=2pt}
		\addlegendentry{BCRB};
		
	\end{axis}
\end{tikzpicture}
		\end{subfigure}\vfill
		\begin{subfigure}{0.48\columnwidth}
			\label{fig_mse_alpha}
			\pgfplotsset{every axis label/.append style={font=\footnotesize},
	every tick label/.append style={font=\footnotesize},
}

\definecolor{clr1}{RGB}{128,0,0}
\definecolor{clr2}{RGB}{170,200,90}
\definecolor{clr5}{RGB}{17,215,205}

\begin{tikzpicture}
	
	\begin{axis}[%
		width=.8\columnwidth,
		height=.6\columnwidth,
		at={(0.758in,0.603in)},
		scale only axis,
		xmin=0,
		xmax=18,
		xlabel style={font=\footnotesize},
		xlabel={SNR (dB)},
		xtick={0,3,6,9,12,15,18},
		xticklabels={0,3,6,9,12,15,18},
		xmajorgrids,
		ymin=1e-4,
		ymax=1e-1,
		yminorticks=true,
		ymode=log,
		ylabel style={at={(axis description cs:-0.13,0.5)},font=\footnotesize},
		ylabel={NMSE of $\bm{\zeta}_\mathrm{real}$},
		ymajorgrids,
		yminorgrids,
		axis background/.style={fill=white},
		]
		
		\addplot [color=green,solid,line width=0.8pt,mark=triangle,mark options={solid},mark size=4pt]
		table[row sep=crcr]{%
			0	0.0868\\ 
			3   0.0521\\
			6   0.0308\\
			9   0.0168\\
			12  0.0142\\
			15  0.0139\\
			18  0.012\\
		};
		

		\addplot [color=blue!50,solid,line width=0.8pt,mark=triangle,mark options={solid},mark size=4pt]
		table[row sep=crcr]{%
			0	0.09\\ 
			3   0.0429\\
			6   0.0176\\
			9   0.0059\\
			12  0.0018\\
			15  7.1063e-04\\
			18  3.4e-04\\
		};
		
		
		
		\addplot [color=clr1!50,solid,line width=0.8pt,mark=triangle,mark options={solid},mark size=4pt]
		table[row sep=crcr]{%
			0	0.036\\ 
			3   0.0119\\
			6   0.0043\\
			9   0.0015\\
			12  7.2e-04\\
			15  3.7e-04\\
			18  1.7998e-04\\
		};

		\addplot [color=black,dashed,line width=0.8pt,mark=triangle,mark options={solid},mark size=4pt]
		table[row sep=crcr]{%
			0	0.0108\\ 
			3   0.0054\\
			6   0.0027\\
			9   0.0014\\
			12  6.8298e-04\\
			15  3.4213e-04\\
			18  1.7274e-04\\
		};
		
		\addplot [color=clr2,solid,line width=0.8pt,mark=triangle,mark options={solid},mark size=4pt]
		table[row sep=crcr]{%
			0	0.0909\\ 
			3   0.031\\
			6   0.0131\\
			9   0.0046\\
			12  0.0019\\
			15  7.6e-04\\
			18  3.8e-04\\
		};
		
		\addplot [color=clr5,solid,line width=0.8pt,mark=triangle,mark options={solid},mark size=4pt]
		table[row sep=crcr]{%
			0	0.0700\\ 
			3   0.032\\
			6   0.0137\\
			9   0.005\\
			12  0.0024\\
			15  0.0017\\ 
			18  0.0016\\
		};
	\end{axis}
	
\end{tikzpicture}%
		\end{subfigure}\hfill
		\begin{subfigure}{0.48\columnwidth}
			\label{fig_ser_ongrid}
			\pgfplotsset{every axis label/.append style={font=\footnotesize},
	every tick label/.append style={font=\footnotesize},
}

\definecolor{clr1}{RGB}{128,0,0}
\definecolor{clr2}{RGB}{170,200,90}
\definecolor{clr5}{RGB}{17,215,205}

\begin{tikzpicture}
	
	\begin{axis}[%
		width=.8\columnwidth,
		height=.6\columnwidth,
		at={(0.758in,0.603in)},
		scale only axis,
		xmin=0,
		xmax=12,
		xlabel style={font=\footnotesize},
		xlabel={SNR (dB)},
		xtick={0,3,6,9,12},
		xticklabels={0,3,6,9,12},
		xmajorgrids,
		ymin=1e-4,
		ymax=1,
		yminorticks=true,
		ymode=log,
		ylabel style={at={(axis description cs:-0.13,0.5)},font=\footnotesize},
		ylabel={BER},
		ymajorgrids,
		yminorgrids,
		axis background/.style={fill=white},
		]
		
		\addplot [color=green,solid,line width=0.8pt,mark=triangle,mark options={solid},mark size=4pt]
		table[row sep=crcr]{%
			0	0.2170\\ 
			3   0.1541\\
			6   0.1026\\
			9   0.0476\\
			12  0.0246\\
		};
		
		
		\addplot [color=blue!50,solid,line width=0.8pt,mark=triangle,mark options={solid},mark size=4pt]
		table[row sep=crcr]{%
			0	0.0845\\ 
			3   0.0427\\
			6   0.0113\\
			9   0.0011\\
			12  2.2396e-04\\
		};
		
		
		\addplot [color=clr1!50,solid,line width=0.8pt,mark=triangle,mark options={solid},mark size=4pt]
		table[row sep=crcr]{%
			0	0.09\\ 
			3   0.013\\
			6   0.0025\\
			9   5e-04\\
			12  1e-4\\
		};
		
		
		
		\addplot [color=clr2,solid,line width=0.8pt,mark=triangle,mark options={solid},mark size=4pt]
		table[row sep=crcr]{%
            0   0.0830\\
			3	0.0360\\ 
			6   0.013\\
			9   0.0015\\
			12  2.2e-4\\
		};
		
		\addplot [color=clr5,solid,line width=0.8pt,mark=triangle,mark options={solid},mark size=4pt]
		table[row sep=crcr]{%
			0   0.0830\\
			3	0.0372\\ 
			6   0.0089\\
			9   0.001\\
			12  2.12e-04\\
		};
	\end{axis}
	
\end{tikzpicture}%
		\end{subfigure}
		\captionof{figure}{The on-grid performance comparison against the SNR.}
		\label{fig_ongrid}
	\end{minipage}
\end{figure}   

Fig. \ref{fig_ongrid} shows the normalized mean square error (NMSE) of $\bm{\zeta}_\mathrm{real}$ and the bit error rate (BER) of the passive information transferred by RISs in the on-grid system, where the localization parameters to be estimated are assumed to be placed on the grid specified in \eqref{equ_grids}, and the indices can be obtained through the positions of non-zero rows in $\bm{\zeta}$. The numbers of grid points in the angle and delay domains are both set to 25. The row-sparse matrix $\bm{\zeta}$ to be recovered is with the size of $625\times20$, and contains 5 non-zero rows, i.e., there are $K=5$ devices in the service region. DQPSK is used for the modulation of passive information of RISs. It is observed that the proposed Algorithm \ref{alg_markov} can approach the BCRB for $\bm{\zeta}_\mathrm{real}$ at high SNR. Moreover, compared with other algorithms listed in Table \ref{tab_complexity}, Algorithm \ref{alg_markov} has the best performance in terms of the NMSE of $\bm{\zeta}_\mathrm{real}$ and the BER of the passive information transferred by RISs. While simulating the BG-GAMP, OMP, Turbo-CS and SBL algorithms, each column of $\bm{\zeta}$ is independently estimated until all the columns are obtained.

We now consider the possible mismatch between the true localization parameters and the parametric model, namely, the off-grid scenario, where $\bm{\theta}_\mathrm{real}$ and $\bm{\tau}_\mathrm{real}$ are randomly selected within their corresponding ranges. The initial grid values are the same with the grids defined in \eqref{equ_grids}. The maximal number of EM-learning iterations is set as 70. In Fig. \ref{equ_crbtuning}, the decreasing NMSE curves of Algorithm \ref{alg1} show that with increasing the number of EM-learning iterations the model mismatch problem is being improved gradually, which verifies the effectiveness of the proposed EM-based parameter learning method. Moreover, Fig. \ref{fig_offgrid} shows the performance of the EM-based parameters learning algorithm against the SNR in the off-grid system. It can be seen that with the proposed learning algorithm both the NMSE and the BER performance significantly outperform those of algorithms without learning methods, especially at high SNR. Note that the proposed EM-based parameters learning method in Algorithm \ref{alg1} can be cooperated with other CS algorithms as long as the posterior mean and variance of $\bm{\zeta}$ can be obtained. Specifically, step 4 in Algorithm \ref{alg1} can be replaced by various CS algorithms. Based on that, we have also presented the performance of BG-GAMP and SBL with the proposed EM-learning method in Fig. \ref{fig_offgrid}, which shows the effectiveness of the proposed learning method as well. 
\begin{figure}[!htbp]
	\begin{minipage}{0.95\columnwidth}
		\begin{subfigure}{0.32\columnwidth}
			\pgfplotsset{every axis label/.append style={font=\footnotesize},
	every tick label/.append style={font=\footnotesize},
}

\definecolor{clr1}{RGB}{128,0,0}
\definecolor{clr3}{RGB}{238,26,196}
\definecolor{clr5}{RGB}{17,215,205}

\begin{tikzpicture}
	
	\begin{axis}[%
		width=.8\columnwidth,
		height=.6\columnwidth,
		at={(0.758in,0.603in)},
		scale only axis,
		xmin=0,
		xmax=70,
		xlabel style={font=\footnotesize},
		xlabel={EM-learning iterations},
		xtick={0,10,20,30,40,50,60,70},
		xticklabels={0,10,20,30,40,50,60,70},
		xmajorgrids,
		ymin=1e-6,
		ymax=1,
		yminorticks=true,
		ymode=log,
		ylabel style={at={(axis description cs:-0.16,0.5)},font=\footnotesize},
		ylabel={NMSE of $\bm{\theta}_{\mathrm{real}}$},
		ymajorgrids,
		yminorgrids,
		axis background/.style={fill=white},
		legend style={at={(0.95,0.9)},anchor=north east,legend cell align=left,align=left,draw=white!15!black,font=\scriptsize}
		]
		
		\addplot [color=clr1,solid,line width=0.8pt,mark=triangle,mark options={solid},mark size=4pt]
		table[row sep=crcr]{%
			0	0.452923170114906\\
			1	0.409994666703189\\
			2	0.294471816286944\\
			3	0.23772153010755\\
			4	0.119602624388424\\
			5	0.166076729294929\\
			6	0.13046291612312\\
			7	0.120083877861613\\
			8	0.132794192709745\\
			9	0.0816179933249773\\
			10	0.0711453432369636\\
			11	0.0711948868377421\\
			12	0.0442340941203151\\
			13	0.0442394013899157\\
			14	0.0712344385831032\\
			15	0.0442398182940662\\
			16	0.0442739356172254\\
			17	0.0441669375271636\\
			18	0.0441806131151074\\
			19	0.0441928403474581\\
			20	0.0442202653759221\\
			21	0.000203777027257267\\
			22	0.000178300414094875\\
			23	0.000157559396085026\\
			24	0.000132017884789812\\
			25	0.00011760254667013\\
			26	9.95967667485466e-05\\
			27	8.38967479725243e-05\\
			28	6.79454250107063e-05\\
			29	5.74805414626891e-05\\
			30	5.12349754810012e-05\\
			31	4.35601835103202e-05\\
			32	3.75644076977282e-05\\
			33	3.10751935460705e-05\\
			34	2.75971038430752e-05\\
			35	2.12592288768035e-05\\
			36	1.95268074048574e-05\\
			37	1.67375938720466e-05\\
			38	1.52701186743136e-05\\
			39	1.32078468137421e-05\\
			40	1.12887450238893e-05\\
			41	1.03951908643148e-05\\
			42	9.85587012469629e-06\\
			43	8.0534727077121e-06\\
			44	7.81323647398365e-06\\
			45	7.18928485720572e-06\\
			46	6.35369039984429e-06\\
			47	6.06446669645443e-06\\
			48	5.48468636009512e-06\\
			49	5.11406717887577e-06\\
			50	4.85736669074837e-06\\
			51	4.38008387201796e-06\\
			52	4.29337066837072e-06\\
			53	4.08830702919202e-06\\
			54	3.82782350863631e-06\\
			55	3.71307125430084e-06\\
			56	3.62556218664653e-06\\
			57	3.31831862560424e-06\\
			58	3.23654227725548e-06\\
			59	3.12181394144094e-06\\
			60	2.98935977996366e-06\\
			61	2.89266443749447e-06\\
			62	2.84719941087548e-06\\
			63	2.82186429383576e-06\\
			64	2.678988721781e-06\\
			65	2.65898441992199e-06\\
			66	2.61082695652695e-06\\
			67	2.50686247940671e-06\\
			68	2.48829789192808e-06\\
			69	2.48166241237617e-06\\
		};	

	\end{axis}
	
\end{tikzpicture}%
		\end{subfigure}\hfil
		\begin{subfigure}{0.32\columnwidth}
			\pgfplotsset{every axis label/.append style={font=\footnotesize},
	every tick label/.append style={font=\footnotesize},
}

\definecolor{clr1}{RGB}{128,0,0}
\definecolor{clr3}{RGB}{238,26,196}
\definecolor{clr5}{RGB}{17,215,205}

\begin{tikzpicture}
	
	\begin{axis}[%
		width=.8\columnwidth,
		height=.6\columnwidth,
		at={(0.758in,0.603in)},
		scale only axis,
		xmin=0,
		xmax=70,
		xlabel style={font=\footnotesize},
		xlabel={EM-learning iterations},
		xtick={0,10,20,30,40,50,60,70},
		xticklabels={0,10,20,30,40,50,60,70},
		xmajorgrids,
		ymin=1e-7,
		ymax=1e-1,
		ymode=log,
		ylabel style={at={(axis description cs:-0.16,0.5)},font=\footnotesize},
		ylabel={NMSE of $\bm{d}_{\mathrm{real}}$},
		ymajorgrids,
		yminorgrids,
		axis background/.style={fill=white},
		legend style={at={(0.95,0.9)},anchor=north east,legend cell align=left,align=left,draw=white!15!black,font=\scriptsize}
		]
		
		\addplot [color=clr1,solid,line width=0.8pt,mark=triangle,mark options={solid},mark size=4pt]
		table[row sep=crcr]{%
			0	0.0401860581528885\\
			1	0.0329716797203289\\
			2	0.0215894092963515\\
			3	0.0198454958722739\\
			4	0.0105141765663498\\
			5	0.0165462426197328\\
			6	0.0122917874393334\\
			7	0.0109482407998626\\
			8	0.0122806548571471\\
			9	0.00718856449474428\\
			10	0.0058479203235181\\
			11	0.00584938380383438\\
			12	0.00237620132034005\\
			13	0.00237317743943673\\
			14	0.00584325541814731\\
			15	0.00237822924890486\\
			16	0.00237811829327239\\
			17	0.00237981431796352\\
			18	0.00237935046930715\\
			19	0.00238283918876918\\
			20	0.00238779598717807\\
			21	9.32254989350076e-06\\
			22	8.5374569495028e-06\\
			23	7.59594335805954e-06\\
			24	6.94157888326498e-06\\
			25	5.93458884308301e-06\\
			26	5.28063613453157e-06\\
			27	4.73680472654842e-06\\
			28	4.41910950926204e-06\\
			29	3.71006169357777e-06\\
			30	3.37736042471773e-06\\
			31	2.93826424139104e-06\\
			32	2.73216459007546e-06\\
			33	2.3376305783577e-06\\
			34	2.21174823116238e-06\\
			35	1.72891577845955e-06\\
			36	1.64992364122874e-06\\
			37	1.41797840464584e-06\\
			38	1.32609564788508e-06\\
			39	1.10382007712785e-06\\
			40	1.04791431889208e-06\\
			41	9.46874092059065e-07\\
			42	8.95065089195782e-07\\
			43	7.70145553602655e-07\\
			44	7.01136864325375e-07\\
			45	6.5012374824473e-07\\
			46	6.10380936623679e-07\\
			47	5.37608194850792e-07\\
			48	5.17063409933529e-07\\
			49	4.58759487174483e-07\\
			50	4.36169878891097e-07\\
			51	4.10000685906662e-07\\
			52	3.88122070674572e-07\\
			53	3.56826275406407e-07\\
			54	3.44137375829887e-07\\
			55	3.27470735518783e-07\\
			56	3.16126395991999e-07\\
			57	2.89232886553367e-07\\
			58	2.80884061797947e-07\\
			59	2.65813878657806e-07\\
			60	2.49036091463245e-07\\
			61	2.35380133871187e-07\\
			62	2.29419467913733e-07\\
			63	2.22454246030839e-07\\
			64	2.18431240087268e-07\\
			65	2.12771579191334e-07\\
			66	1.98327580610654e-07\\
			67	1.89728571557934e-07\\
			68	1.86792599491169e-07\\
			69	1.82469573164005e-07\\
		};	
	\end{axis}
	
\end{tikzpicture}%
		\end{subfigure}\hfil
		\begin{subfigure}{0.32\columnwidth}
			\pgfplotsset{every axis label/.append style={font=\footnotesize},
	every tick label/.append style={font=\footnotesize},
}

\definecolor{clr1}{RGB}{128,0,0}
\definecolor{clr3}{RGB}{238,26,196}
\definecolor{clr5}{RGB}{17,215,205}

\begin{tikzpicture}
	
	\begin{axis}[%
		width=.8\columnwidth,
		height=.6\columnwidth,
		at={(0.758in,0.603in)},
		scale only axis,
		xmin=0,
		xmax=70,
		xlabel style={font=\footnotesize},
		xlabel={EM-learning iterations},
		xtick={0,10,20,30,40,50,60,70},
		xticklabels={0,10,20,30,40,50,60,70},
		xmajorgrids,
		ymin=1e-4,
		ymax=1,
		ymode=log,
		ylabel style={at={(axis description cs:-0.16,0.5)},font=\footnotesize},
		ylabel={NMSE of $\bm{\zeta}_{\mathrm{real}}$},
		ytick={1e-4,1e-3,1e-2,1e-1,1},
		yticklabels={$10^{-4}$,$10^{-3}$,$10^{-2}$,$10^{-1}$,$10^{0}$},
		ymajorgrids,
		yminorgrids,
		axis background/.style={fill=white},
		legend style={at={(0.95,0.9)},anchor=north east,legend cell align=left,align=left,draw=white!15!black,font=\scriptsize}
		]
		
		\addplot [color=clr1,solid,line width=0.8pt,mark=triangle,mark options={solid},mark size=4pt]
		table[row sep=crcr]{%
			0	0.403899118115459\\
			1	0.310799914624652\\
			2	0.243053620811283\\
			3	0.195113177120128\\
			4	0.159188042375156\\
			5	0.128344660199007\\
			6	0.104445736281623\\
			7	0.0899373427998654\\
			8	0.0741115584080779\\
			9	0.0604134657148966\\
			10	0.0515254929559476\\
			11	0.042312029854034\\
			12	0.0324219376864111\\
			13	0.0261471278636329\\
			14	0.024821111717793\\
			15	0.0181566567679678\\
			16	0.0150899998832817\\
			17	0.0126236682056178\\
			18	0.0108569645251842\\
			19	0.00925608687200257\\
			20	0.0078177796181242\\
			21	0.00652402331650896\\
			22	0.0053977012951734\\
			23	0.00464619250604002\\
			24	0.00364866998109036\\
			25	0.00320329339372155\\
			26	0.00256862001337008\\
			27	0.00207807002527707\\
			28	0.00167610194867874\\
			29	0.00137474727877209\\
			30	0.00120794370930083\\
			31	0.00104761075258993\\
			32	0.000931581761895567\\
			33	0.000805540297994848\\
			34	0.000749453809895791\\
			35	0.000624569948818132\\
			36	0.000592322218918745\\
			37	0.000528327719574449\\
			38	0.000503171043215457\\
			39	0.00045428539043377\\
			40	0.000429839327379411\\
			41	0.00040652725172019\\
			42	0.000394748632143799\\
			43	0.000364683130266829\\
			44	0.000354998936145511\\
			45	0.000343265202174226\\
			46	0.00033065521696815\\
			47	0.000321712597860058\\
			48	0.000313179720003576\\
			49	0.000306133603183618\\
			50	0.000298819235407243\\
			51	0.000291735350219914\\
			52	0.000287676956715261\\
			53	0.000284279974050168\\
			54	0.000281310151127597\\
			55	0.000277819515890507\\
			56	0.000275409112546394\\
			57	0.000270408165887106\\
			58	0.000269223139801851\\
			59	0.00026725174611605\\
			60	0.000264355317079233\\
			61	0.000262205804496747\\
			62	0.000261381159361672\\
			63	0.000260201198754357\\
			64	0.000259721327951119\\
			65	0.000258517592498084\\
			66	0.000256678768957433\\
			67	0.000254608157789673\\
			68	0.000254506101425957\\
			69	0.000254703565716048\\
		};	
\end{axis}

\end{tikzpicture}%
		\end{subfigure}
		\caption{The off-grid NMSE of Algorithm \ref{alg1} against the number of EM-learning iterations in the angle $\bm{\theta}_{\mathrm{real}}$, distance $\bm{d}_{\mathrm{real}}$ and $\bm{\zeta}_{\text{real}}$ domain under $\text{SNR}=18 \text{ dB}$.}
		\label{equ_crbtuning}
	\end{minipage}
\end{figure}
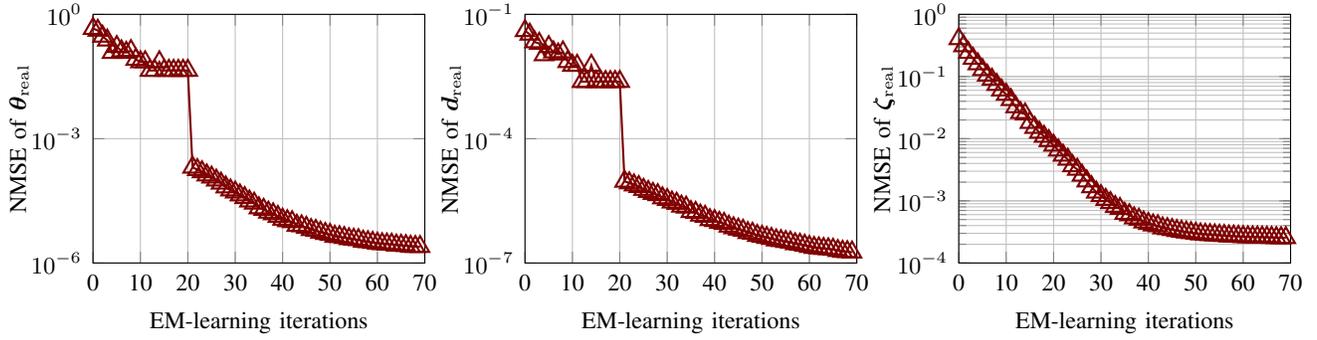
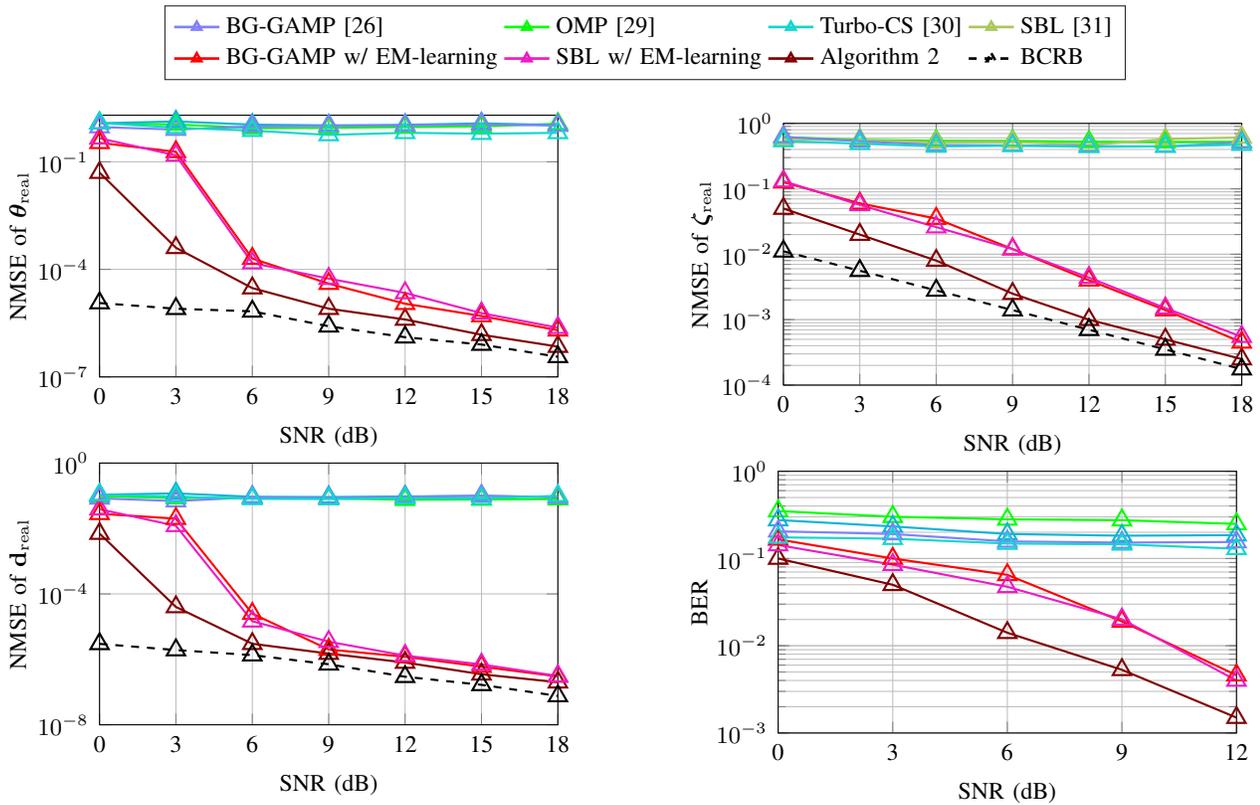
\begin{figure}[!htbp]
	\begin{minipage}{0.95\columnwidth}
		\hspace{2.2cm}\begin{subfigure}{0.95\columnwidth}
			\definecolor{clr2}{RGB}{170,200,90}
\definecolor{clr3}{RGB}{238,26,196}
\definecolor{clr1}{RGB}{128,0,0}
\definecolor{clr5}{RGB}{17,215,205}
\definecolor{clr6}{RGB}{17,215,205}

\begin{tikzpicture} 
	\begin{axis}[%
		hide axis,
		xmin=10,
		xmax=40,
		ymin=0,
		ymax=0.4,
		legend columns=4,
		legend style={draw=white!15!black,legend cell align=left,font=\footnotesize}
		]
		
		\addlegendimage{blue!50,line width=1.0pt,mark=triangle,mark size=2pt}
		\addlegendentry{BG-GAMP \cite{6556987}};
		
		\addlegendimage{green,line width=1.0pt,mark=triangle,mark size=2pt}
		\addlegendentry{OMP \cite{4385788}};
		
		\addlegendimage{clr5,line width=1.0pt,mark=triangle,mark size=2pt}
		\addlegendentry{Turbo-CS \cite{6883198}};
		
		\addlegendimage{clr2,line width=1.0pt,mark=triangle,mark size=2pt}
		\addlegendentry{SBL \cite{tipping2001sparse}};
		
		\addlegendimage{red,line width=1.0pt,mark=triangle,mark size=2pt}
		\addlegendentry{BG-GAMP w/ EM-learning};
		
		
		\addlegendimage{clr3,line width=1.0pt,mark=triangle,mark size=2pt}
		\addlegendentry{SBL w/ EM-learning};
		
		\addlegendimage{clr1,line width=1.0pt,mark=triangle,mark size=2pt}
		\addlegendentry{Algorithm \ref{alg1}};
		
		\addlegendimage{black,dashed,line width=1.0pt,mark=triangle,mark size=2pt}
		\addlegendentry{BCRB};
		
	\end{axis}
\end{tikzpicture}
		\end{subfigure}\vspace{0.3cm}
	\end{minipage}\vfill
	\begin{minipage}{0.48\columnwidth}
		\begin{subfigure}{\columnwidth}
			\pgfplotsset{every axis label/.append style={font=\footnotesize},
	every tick label/.append style={font=\footnotesize},
}

\definecolor{clr2}{RGB}{13,180,215}
\definecolor{clr1}{RGB}{128,0,0}
\definecolor{clr3}{RGB}{238,26,196}
\definecolor{clr5}{RGB}{17,215,205}

\begin{tikzpicture}
	
	\begin{axis}[%
		width=.7\columnwidth,
		height=.4\columnwidth,
		at={(0.758in,0.603in)},
		scale only axis,
		xmin=0,
		xmax=18,
		xlabel style={font=\footnotesize},
		xlabel={SNR (dB)},
		xtick={0,3,6,9,12,15,18},
		xticklabels={0,3,6,9,12,15,18},
		xmajorgrids,
		ymin=1e-7,
		ymax=2,
		yminorticks=true,
		ymode=log,
		ylabel style={at={(axis description cs:-0.13,0.5)},font=\footnotesize},
		ylabel={NMSE of $\bm{\theta}_\mathrm{real}$},
		ymajorgrids,
		yminorgrids,
		axis background/.style={fill=white},
		]
		
		\addplot [color=green,solid,line width=0.8pt,mark=triangle,mark options={solid},mark size=4pt]
		table[row sep=crcr]{%
			0	1.1952\\ 
			3   1.1008\\
			6	0.8767\\
			9   0.8850\\
			12  0.9332\\
			15  0.9748\\
			18  1.1528\\
		};
				
		
		\addplot [color=clr2,solid,line width=0.8pt,mark=triangle,mark options={solid},mark size=4pt]
		table[row sep=crcr]{%
			0	1.1904\\ 
			3   1.3383\\
			6	1.0894\\
			9   1.0415\\
			12  1.0761\\
			15  1.1853\\
			18  1.0623\\
		};
		
		
		
		\addplot [color=clr1,solid,line width=0.8pt,mark=triangle,mark options={solid},mark size=4pt]
		table[row sep=crcr]{%
			0  0.05\\
			3  4e-04\\
			6  3e-05\\
			9  8e-6\\  
			12  4e-6\\
			15  1.5e-6\\
			18  7e-7\\
		};
		
		\addplot [color=blue!50,solid,line width=0.8pt,mark=triangle,mark options={solid},mark size=4pt]
		table[row sep=crcr]{%
			0	0.9336\\ 
			3   0.7944\\
			6	0.9742\\
			9   1.0006\\
			12  1.0324\\
			15  1.1153\\
			18  1.0252\\
		};
		
		\addplot [color=red,solid,line width=0.8pt,mark=triangle,mark options={solid},mark size=4pt]
		table[row sep=crcr]{%
			0	0.3399\\ 
			3   0.19\\
			6	2e-4\\
			9   4e-5\\
			12  1.1e-5\\
			15  5e-06\\
			18  2e-06\\
		};
		
		\addplot [color=black,dashed,line width=0.8pt,mark=triangle,mark options={solid},mark size=4pt]
		table[row sep=crcr]{%
			0	1.1508e-05\\ 
			3   7.9751e-06\\
			6	6.7459e-06\\
			9   2.5868e-06\\
			12  1.2811e-06\\
			15  7.9981e-07\\
			18  3.6378e-07\\
		};
		
		
		\addplot [color=clr3,solid,line width=0.8pt,mark=triangle,mark  options={solid},mark size=4pt]
		table[row sep=crcr]{%
			0	0.4540\\ 
			3   0.1515\\
			6	1.5e-04\\
			9   5.5164e-05\\
			12  2.2e-05\\
			15  6e-06\\
			18  2.3637e-06\\
		};
		
		\addplot [color=clr5,solid,line width=0.8pt,mark=triangle,mark options={solid},mark size=4pt]
		table[row sep=crcr]{%
			0	1.2115\\ 
			3   0.8967\\
			6	0.7413\\
			9   0.5628\\
			12  0.6453\\
			15  0.6103\\
			18  0.6423\\
		};
	\end{axis}
	
\end{tikzpicture}%
		\end{subfigure}\vspace{-0.2cm}
		\begin{subfigure}{\columnwidth}
			\pgfplotsset{every axis label/.append style={font=\footnotesize},
	every tick label/.append style={font=\footnotesize},
}

\definecolor{clr2}{RGB}{13,180,215}
\definecolor{clr1}{RGB}{128,0,0}
\definecolor{clr3}{RGB}{238,26,196}
\definecolor{clr5}{RGB}{17,215,205}

\begin{tikzpicture}
	
	\begin{axis}[%
		width=.7\columnwidth,
		height=.4\columnwidth,
		at={(0.758in,0.603in)},
		scale only axis,
		xmin=0,
		xmax=18,
		xlabel style={font=\footnotesize},
		xlabel={SNR (dB)},
		xtick={0,3,6,9,12,15,18},
		xticklabels={0,3,6,9,12,15,18},
		xmajorgrids,
		ymin=1e-8,
		ymax=1,
		yminorticks=true,
		ymode=log,
		ylabel style={at={(axis description cs:-0.13,0.5)},font=\footnotesize},
		ylabel={NMSE of $\mathbf{d}_\mathrm{real}$},
		ymajorgrids,
		yminorgrids,
		axis background/.style={fill=white},
		]
		
		\addplot [color=green,solid,line width=0.8pt,mark=triangle,mark options={solid},mark size=4pt]
		table[row sep=crcr]{%
			0	0.0968\\ 
			3   0.0856\\
			6	0.0855\\
			9   0.0832\\
			12  0.0758\\
			15  0.0771\\
			18  0.0793\\
		};
		
		
		\addplot [color=clr2,solid,line width=0.8pt,mark=triangle,mark options={solid},mark size=4pt]
		table[row sep=crcr]{%
			0	0.1076\\ 
			3   0.1185\\
			6	0.0932\\
			9   0.0916\\
			12  0.0956\\
			15  0.0982\\
			18  0.0929\\
		};
		
		
		
		\addplot [color=clr1,solid,line width=0.8pt,mark=triangle,mark options={solid},mark size=4pt]
		table[row sep=crcr]{%
			0  0.007\\
			3  4e-5\\
			6  3e-6\\
			9  1.5e-6\\
			12  8e-7\\
			15  3.5e-07\\
			18  2e-07\\
		};
		
		\addplot [color=blue!50,solid,line width=0.8pt,mark=triangle,mark options={solid},mark size=4pt]
		table[row sep=crcr]{%
			0	0.0836\\ 
			3   0.0690\\
			6	0.0898\\
			9   0.0848\\
			12  0.0932\\
			15  0.1012\\
			18  0.0888\\
		};
		
		\addplot [color=red,solid,line width=0.8pt,mark=triangle,mark options={solid},mark size=4pt]
		table[row sep=crcr]{%
			0	0.0282\\ 
			3   0.02\\
			6	2.5e-05\\
			9   2e-06\\
			12  1.2e-06\\
			15  6e-07\\
			18  3e-07\\
		};
		
		\addplot [color=black,dashed,line width=0.8pt,mark=triangle,mark options={solid},mark size=4pt]
		table[row sep=crcr]{%
			0	2.9770e-06\\ 
			3   1.9270e-06\\
			6	1.3454e-06\\
			9  7.0782e-07\\
			12  2.9449e-07\\
			15  1.6641e-07\\
			18  7.6075e-08\\
		};
		
		
		\addplot [color=clr3,solid,line width=0.8pt,mark=triangle,mark  options={solid},mark size=4pt]
		table[row sep=crcr]{%
			0	0.0389\\ 
			3   0.0121\\
			6	1.4624e-05\\
			9   3.4650e-06\\
			12  1.3e-6\\
			15  7e-07\\ 
			18  3.1e-07\\  
		};
		
		\addplot [color=clr5,solid,line width=0.8pt,mark=triangle,mark options={solid},mark size=4pt]
		table[row sep=crcr]{%
			0	0.1073\\ 
			3   0.0948\\
			6	0.0805\\
			9   0.0796\\
			12  0.0846\\
			15  0.0821\\
			18  0.1003\\
		};
	\end{axis}
	
\end{tikzpicture}%
		\end{subfigure}
	\end{minipage}\hfil
	\begin{minipage}{0.48\columnwidth}
		\begin{subfigure}{\columnwidth}
			\pgfplotsset{every axis label/.append style={font=\footnotesize},
	every tick label/.append style={font=\footnotesize},
}

\definecolor{clr3}{RGB}{238,26,196}
\definecolor{clr2}{RGB}{170,200,90}
\definecolor{clr1}{RGB}{128,0,0}
\definecolor{clr5}{RGB}{17,215,205}

\begin{tikzpicture}
	
	\begin{axis}[%
		width=.7\columnwidth,
		height=.4\columnwidth,
		at={(0.758in,0.603in)},
		scale only axis,
		xmin=0,
		xmax=18,
		xlabel style={font=\footnotesize},
		xlabel={SNR (dB)},
		xtick={0,3,6,9,12,15,18},
		xticklabels={0,3,6,9,12,15,18},
		xmajorgrids,
		ymin=1e-4,
		ymax=1,
		yminorticks=true,
		ymode=log,
		ylabel style={at={(axis description cs:-0.13,0.5)},font=\footnotesize},
		ylabel={NMSE of $\bm{\zeta}_\mathrm{real}$},
		ytick={1e-4,1e-3,1e-2,1e-1,1e0},
		yticklabels={$10^{-4}$,$10^{-3}$,$10^{-2}$,$10^{-1}$,$10^{0}$},
		ymajorgrids,
		yminorgrids,
		axis background/.style={fill=white},
		]
		
		\addplot [color=green,solid,line width=0.8pt,mark=triangle,mark options={solid},mark size=4pt]
		table[row sep=crcr]{%
			0	0.6176\\ 
			3   0.5597\\
			6	0.5429\\
			9   0.5404\\
			12  0.5243\\
			15  0.5209\\
			18  0.5239\\
		};
		

	\addplot [color=clr2,solid,line width=0.8pt,mark=triangle,mark options={solid},mark size=4pt]
	table[row sep=crcr]{%
		0	0.5768\\ 
		3   0.5494\\
		6	0.5180\\
		9   0.5218\\
		12  0.4665\\
		15  0.5753\\
		18  0.6100\\
	};

		
		\addplot [color=clr1,solid,line width=0.8pt,mark=triangle,mark options={solid},mark size=4pt]
		table[row sep=crcr]{%
			0  0.05\\
			3  0.02\\
			6  0.008\\
			9  0.0025\\
			12  1e-3\\
			15  5e-04\\
			18  2.5e-04\\
		};
		
		\addplot [color=blue!50,solid,line width=0.8pt,mark=triangle,mark options={solid},mark size=4pt]
		table[row sep=crcr]{%
			0	0.6227\\ 
			3   0.5341\\
			6	0.4668\\
			9   0.4612\\
			12  0.4502\\
			15  0.4443\\
			18  0.5119\\
		};
		
		\addplot [color=red,solid,line width=0.8pt,mark=triangle,mark options={solid},mark size=4pt]
		table[row sep=crcr]{%
			0	0.1261\\ 
			3   0.06\\
			6	0.0351\\
			9   0.012\\
			12  0.004\\
			15  0.0014\\
			18  4.5022e-04\\
		};
		
		\addplot [color=black,dashed,line width=0.8pt,mark=triangle,mark options={solid},mark size=4pt]
		table[row sep=crcr]{%
			0	0.0111\\ 
			3   0.0056\\
			6	0.0028\\
			9   0.0014\\
			12  7.0103e-04\\
			15  3.5062e-04\\
			18  1.7533e-04\\
		};
		
		
		
		\addplot [color=clr3,solid,line width=0.8pt,mark=triangle,mark  options={solid},mark size=4pt]
		table[row sep=crcr]{%
			0	0.1311\\ 
			3   0.0568\\
			6	0.026\\
			9   0.0120\\
			12  0.0044\\
			15  0.0015\\
			18  5.4988e-04\\
		};
		
		\addplot [color=clr5,solid,line width=0.8pt,mark=triangle,mark options={solid},mark size=4pt]
		table[row sep=crcr]{%
			0	0.5390\\ 
			3   0.4903\\
			6	0.4437\\
			9   0.4550\\
			12  0.4380\\
			15  0.4445\\
			18  0.4731\\
		};
	\end{axis}
	
\end{tikzpicture}%
		\end{subfigure}\vspace{-0.2cm}
		\begin{subfigure}{\columnwidth}
			\pgfplotsset{every axis label/.append style={font=\footnotesize},
	every tick label/.append style={font=\footnotesize},
}

\definecolor{clr2}{RGB}{13,180,215}
\definecolor{clr1}{RGB}{128,0,0}
\definecolor{clr3}{RGB}{238,26,196}
\definecolor{clr5}{RGB}{17,215,205}

\begin{tikzpicture}
	
	\begin{axis}[%
		width=.7\columnwidth,
		height=.4\columnwidth,
		at={(0.758in,0.603in)},
		scale only axis,
		xmin=0,
		xmax=12,
		xlabel style={font=\footnotesize},
		xlabel={SNR (dB)},
		xtick={0,3,6,9,12},
		xticklabels={0,3,6,9,12},
		xmajorgrids,
		ymin=1e-3,
		ymax=1,
		yminorticks=true,
		ymode=log,
		ylabel style={at={(axis description cs:-0.13,0.5)},font=\footnotesize},
		ylabel={BER},
		ymajorgrids,
		yminorgrids,
		axis background/.style={fill=white},
		]
		
		\addplot [color=green,solid,line width=0.8pt,mark=triangle,mark options={solid},mark size=4pt]
		table[row sep=crcr]{%
			0	0.3501\\ 
			3   0.3013\\
			6	0.2813\\
			9   0.2751\\
			12  0.2492\\
			15  0.2443\\
			18  0.2414\\
		};
		

		\addplot [color=clr2,solid,line width=0.8pt,mark=triangle,mark options={solid},mark size=4pt]
		table[row sep=crcr]{%
			0	0.2760\\ 
			3   0.2332\\
			6	0.1910\\
			9   0.1827\\
			12  0.1854\\
			15  0.1848\\
			18  0.1547\\
		};
		
		
		
		\addplot [color=clr1,solid,line width=0.8pt,mark=triangle,mark options={solid},mark size=4pt]
		table[row sep=crcr]{%
			0	0.1\\ 
			3   0.05\\
			6	0.0141\\
			9   0.0053\\
			12  0.0015\\
		};
		
		\addplot [color=blue!50,solid,line width=0.8pt,mark=triangle,mark options={solid},mark size=4pt]
		table[row sep=crcr]{%
			0	0.2053\\ 
			3   0.1909\\
			6	0.1570\\
			9   0.1523\\
			12  0.1546\\
			15  0.1638\\
			18  0.1284\\
		};
		
		\addplot [color=red,solid,line width=0.8pt,mark=triangle,mark options={solid},mark size=4pt]
		table[row sep=crcr]{%
			0	0.1658\\ 
			3   0.1\\
			6	0.065\\
			9   0.019\\
			12  0.0046\\
		};
		
		
		\addplot [color=clr3,solid,line width=0.8pt,mark=triangle,mark  options={solid},mark size=4pt]
		table[row sep=crcr]{%
			0	0.1428\\ 
			3   0.0849\\
			6	0.0474\\
			9   0.0201\\
			12  0.004\\ 
		};
		
		\addplot [color=clr5,solid,line width=0.8pt,mark=triangle,mark options={solid},mark size=4pt]
		table[row sep=crcr]{%
			0	0.1749\\ 
			3   0.1688\\
			6	0.1485\\
			9   0.1453\\
			12  0.1295\\
			15  0.1224\\
			18  0.1560\\ 
		};
	\end{axis}
	
\end{tikzpicture}%
		\end{subfigure}
	\end{minipage}
	\caption{The off-grid NMSE performance comparison against the SNR.}
	\label{fig_offgrid}
\end{figure}

\section{Conclusions}\label{sec6}
In this work, we have proposed a new RIS-aided integrated sensing and communication scenario, where a BS communicates with multiple devices in full-duplex, and senses the positions of these devices simultaneously. A grid based parametric model was constructed, and the joint estimation problem was formulated as a CS problem. A novel message-passing algorithm was shown to be used to solve the problem, and the progressive approximation method was proposed to reduce the computational complexity during the message passing. To tackle the issue of model mismatch, the EM algorithm was utilized for the parameters learning. Simulation results have shown good performance of the proposed method.

\bibliographystyle{IEEEtran}
\bibliography{ref}

\end{document}